\documentclass[%
 reprint, superscriptaddress,
 amsmath,amssymb,
 aps,
]{revtex4-2}
\usepackage{soul}
\usepackage{xcolor}
\usepackage{float}
\definecolor{softcyan}{RGB}{200, 240, 250}
\sethlcolor{softcyan}
\usepackage{graphicx}
\usepackage{dcolumn}
\usepackage{bm}
\usepackage{amsmath}
\usepackage{xcolor}
\usepackage[most]{tcolorbox}
\tcbuselibrary{theorems}
\usepackage{empheq}
\usepackage{tikz}
\definecolor{pink}{rgb}{0.78,0.08,0.52}
\usepackage{hyperref}
\hypersetup{
    colorlinks=true,
    linkcolor=magenta,
    citecolor=blue,
    filecolor=magenta,
    urlcolor=violet,
}
\definecolor{DarkBlue}{RGB}{0, 32, 96}

\begin{document}

\title{Postselection-induced localization and coherence in  quantum walks on
heterogeneous networks}

\author{Adithya L J}
\email{adithya.lj@students.iiserpune.ac.in}
\affiliation{Department of Physics, Indian Institute of Science Education and Research Pune, 411008, India}

\author{Suraj S Hegde}
\email{surajhegde@iisertvm.ac.in}
\affiliation{School of Physics, Indian Institute of Science Education and Research
Thiruvananthapuram, 695551, India}

\author{Chandrakala Meena}
\email{chandrakala@iiserpune.ac.in}
\affiliation{Department of Physics, Indian Institute of Science Education and Research Pune, 411008, India}

\date{\today}

\begin{abstract}

Postselection of quantum trajectories is known to effectively introduce nonlinearity into dynamics of open quantum systems. We study the effect of such nonlinearity in continuous-time quantum walks (CTQWs) on networks with homogeneous and heterogeneous degree distributions. Using the recently proposed nonlinear Lindblad master equation (NLME), we investigate the dynamics under two decoherence mechanisms: Haken–Strobl decoherence and quantum stochastic walk (QSW). Our analysis reveals a striking dichotomy: under Haken–Strobl decoherence the nonlinear contributions precisely cancel, yielding a uniform steady state independent of postselection details under imperfect detection. In stark contrast, QSW decoherence permits postselection in this regime to break dynamical balance on heterogeneous networks, inducing robust localization preferentially at low-degree (peripheral) nodes. Remarkably, this localized state maintains finite quantum coherence. Extending our results to spin systems, we demonstrate that degree heterogeneity similarly stabilizes localization of spin-up excitations in spin-down backgrounds, enhancing pairwise entanglement preservation. These findings establish degree heterogeneity and postselection as joint control parameters for engineering quantum transport and localization in dissipative systems.

\end{abstract}

\maketitle

\section{Introduction}
Quantum walks (QWs), the quantum counterparts of classical random walks, provide a versatile framework for exploring a wide range of quantum phenomena, including quantum computation to coherent transport in complex systems~\cite{farhi1998quantum,aharonov1993quantum,childs2009universal,venegas2012quantum,jayakody2023revisiting}. Continuous-time quantum walks (CTQWs), in particular, offer a fundamental approach to quantum transport on networks, where quantum coherence and interference enable more efficient exploration of network structures compared to classical dynamics~\cite{childs2004spatial,mulken2011continuous, acharyya2026comprehensive}.
Beyond their applications in quantum search and information processing~\cite{childs2004spatial,chakraborty2025continuous}, CTQWs serve as effective models for diverse physical phenomena, including quantum state transfer in superconducting qubits~\cite{strauch2008theoretical,strauch2009reexamination}, hole dynamics in degenerate spin environments~\cite{carlstrom2016quantum}, and exciton energy transport in biological light-harvesting systems~\cite{plenio2008dephasing,mulken2011continuous}. Consequently, physical realizations of CTQWs have been achieved in different experimental platforms, including quantum computing architectures~\cite{portugal2022implementation,chen2024implementation,portugal2025efficient}, integrated photonic chips~\cite{tang2018experimental}, and coupled waveguide arrays~\cite{wang2020experimental,qu2022experimental,tang2024simulating}. However, any realistic implementation is inevitably subject to decoherence from the system's interaction with its environment, which typically drives the system towards classical behavior~\cite{breuer2002theory,bressanini2022decoherence,stabilityctqw}.

 The dynamics of open quantum systems under decoherence are comprehensively described by the Gorini-Kossakowski-Sudarshan-Lindblad (GKSL) master equation~\cite{lindblad1976generators}, hereafter referred to as the linear master equation (LME). From the quantum trajectory perspective, the LME maintains dynamical balance through unconditional ensemble averaging of non-Hermitian Hamiltonian evolution interrupted by random quantum jumps. However, postselection, where measured outcomes are conditioned on specific detection events, breaks this averaging procedure and reveals the non-Hermitian physics inherent to normalized conditional states. This motivated the recent introduction of a nonlinear Lindblad master equation (NLME) by Liu and Chen~\cite{liu2025lindbladian}, which provides a systematic description of open quantum systems subject to measurement and postselection.
The NLME reveals that selectively conditioning on measurement outcomes can fundamentally alter system dynamics, giving rise to striking non-trivial phenomena. A notable example is the postselected skin effect~\cite{liu2025lindbladian}, wherein fermions are preferentially driven to one edge of a linear chain. Postselecting for no-jump trajectories restricts the evolution to paths where no quantum jumps occur, providing a well-established method for probing non-Hermitian physics in open quantum systems. By eliminating measurement outcomes associated with decoherence events, this approach exposes the underlying non-Hermitian Hamiltonian dynamics that would otherwise be obscured by the averaging inherent in the open system description~\cite{lee2014heralded,yamamoto2019theory,xu2020topological,liu2020non,yang2021exceptional,chaduteau2025lindbladian,zeng2025non}. Recent experiments have also demonstrated the ability to steer effective non-Hermitian dynamics through postselection on specific quantum jump trajectories in a superconducting transmon circuit~\cite{naghiloo2019quantum}. Indeed, this approach is emerging as a powerful tool for quantum state engineering; for instance, the NLME~\cite{liu2025lindbladian} has been proposed as a method to enhance the generation of tripartite entanglement in spin-qubit systems by steering the dynamics towards a robust, highly entangled dark state~\cite{driessen2025robust}. This method will also be particularly impactful in experimental settings including superconducting qubit systems ~\cite{naghiloo2019quantum,wu2019observation} and optical systems~\cite{wang2023non,liu2025lindbladian}. 

In this work, we investigate how postselection and the resulting nonlinearity influences the interplay between the network topology and the nature of the decoherence models in CTQWs. We also extend the QSW analysis to spin networks to investigate whether network-topological effects persist in the presence of quantum entanglement, specifically by studying single-spin excitation transport. 

This paper is organized as follows. In Sec.~\ref{theory}, we establish the theoretical framework and decoherence models used in our study. In Sec.~\ref{sec:QSW}, we present our main findings, focusing on the interplay between network topology and nonlinearity, which results in unique steady-states under QSW decoherence. We provide an analytical constraint for the steady state and numerically investigate how the network topology and control parameters govern the resulting steady states. Subsequently, we contrast these results with the steady states obtained under Haken-Strobl model. In the Sec.~\ref{spinnetwork}, we extend our analysis to spin networks, examining the steady state and pairwise concurrence in spin networks to determine the fate of quantum entanglement. We conclude the main findings of our study in Sec.~\ref{conclusion} with a summary and outlook.

\section{Theoretical Framework}
\label{theory}
\subsection{Nonlinear Lindblad Master Equation}

The unconditional dynamics of open quantum systems are described by the LME, which models the evolution of the system's density matrix $\rho$. A solution of the LME can be written as the ensemble average over pure-state stochastic evolutions (quantum trajectories) corresponding to particular measurements of the environment. On the other hand, modern experimental platforms, such as photonic systems, allow for the monitoring of individual quantum trajectories conditioned on specific measurement records i.e., postselection~\cite{murch2013observing,weber2014mapping,guerlin2007progressive,rossi2019observing}. Motivated by this, we consider the conditional evolution of the system continuously monitored by imperfect detectors with finite efficiency $\eta \in [0,1)$, and discard the jumps that are detected (postselection).

 For a perfect detector ($\eta=1$), postselection leads to a pure non-Hermitian evolution by discarding all the trajectories with jumps; however under finite efficiency ($\eta<1$) considered here, undetected jumps are retained in the postselected dynamics. Crucially, to describe the physical state conditioned on no discarded jumps, the density matrix must be continuously renormalized to satisfy $\mathrm{Tr}[\rho(t)]=1$. This physical requirement introduces a nonlinearity into the dynamical equation. The resulting normalized conditional evolution of the system's density matrix $\rho$ discarding detected jumps is governed by the NLME~\cite{liu2025lindbladian}:

\begin{equation}
\begin{split}
\frac{d\rho}{dt} = -i[H,\rho]
+ \sum_{k=1}^{M} \gamma \Big(
& -\tfrac{1}{2}\{P_{k}^{\dagger}P_{k}, \rho\} + (1-\eta)\,P_{k}\rho P_{k}^{\dagger}\\
&
+ \eta\,\mathrm{Tr}\!\left(P_{k}^{\dagger}P_{k}\rho\right)\rho
\Big).
\end{split}
\label{eq:1}
\end{equation}

Here, $H$ is the system Hamiltonian, and $\{P_{k}\}$ are the jump operators acting with rates $\gamma$. The first two terms in the summation represent the standard dissipative evolution modified by the detection efficiency. The third term is the nonlinear feedback gain arising from the renormalization. This term depends on the instantaneous state of the system via the trace, introducing a state dependent feedback that breaks the detailed balance of the standard master equation~\cite{liu2025lindbladian}. In the absence of postselection ($\eta=0$), the NLME reduces to standard LME.

\subsection{Continuous-time Quantum Walks on Networks}

We apply the NLME formalism to CTQWs defined on a graph $G=(V,E)$. The coherent dynamics are governed by the graph Laplacian, also termed the Hamiltonian, $H = D - A$. The matrix elements of $H$ are given by:

\begin{equation}
H_{kj} =
\begin{cases}
D_{k}, & \text{if } j = k, \\
-A_{kj}, & \text{if } j \neq k .
\end{cases}
\label{eq:13}
\end{equation}
where $D_{k}$ is the degree or number of connections of $k^{th}$ node and $A_{kj}=1$ if nodes $k$ and $j$ are connected; otherwise, it is zero. This Laplacian based Hamiltonian has been successfully implemented in various experimental platforms, including photonic waveguide arrays~\cite{qu2022experimental} and, more recently, digital quantum circuits~\cite{chakraborty2025continuous}.

\subsection{Decoherence Models}
The jump operators in general lead to either pure dephasing or decoherence of the system's density matrix during the evolution. We are interested in investigating these mechanisms within the context of quantum walks on networks  governed by the NLME. Our primary motivation is to determine how postselection modifies the dynamics under specific decoherence models, steering the system toward steady states that cannot be attained under standard LME. We investigate the interplay between network topology and dissipative dynamics by analyzing two distinct decoherence models, namely quantum stochastic walk (QSW) and Haken-Strobl. These models are distinguished by the spatial nature of their jump operators. 

\begin{figure}[htbp]
\centering
\includegraphics[width=0.36\textwidth]{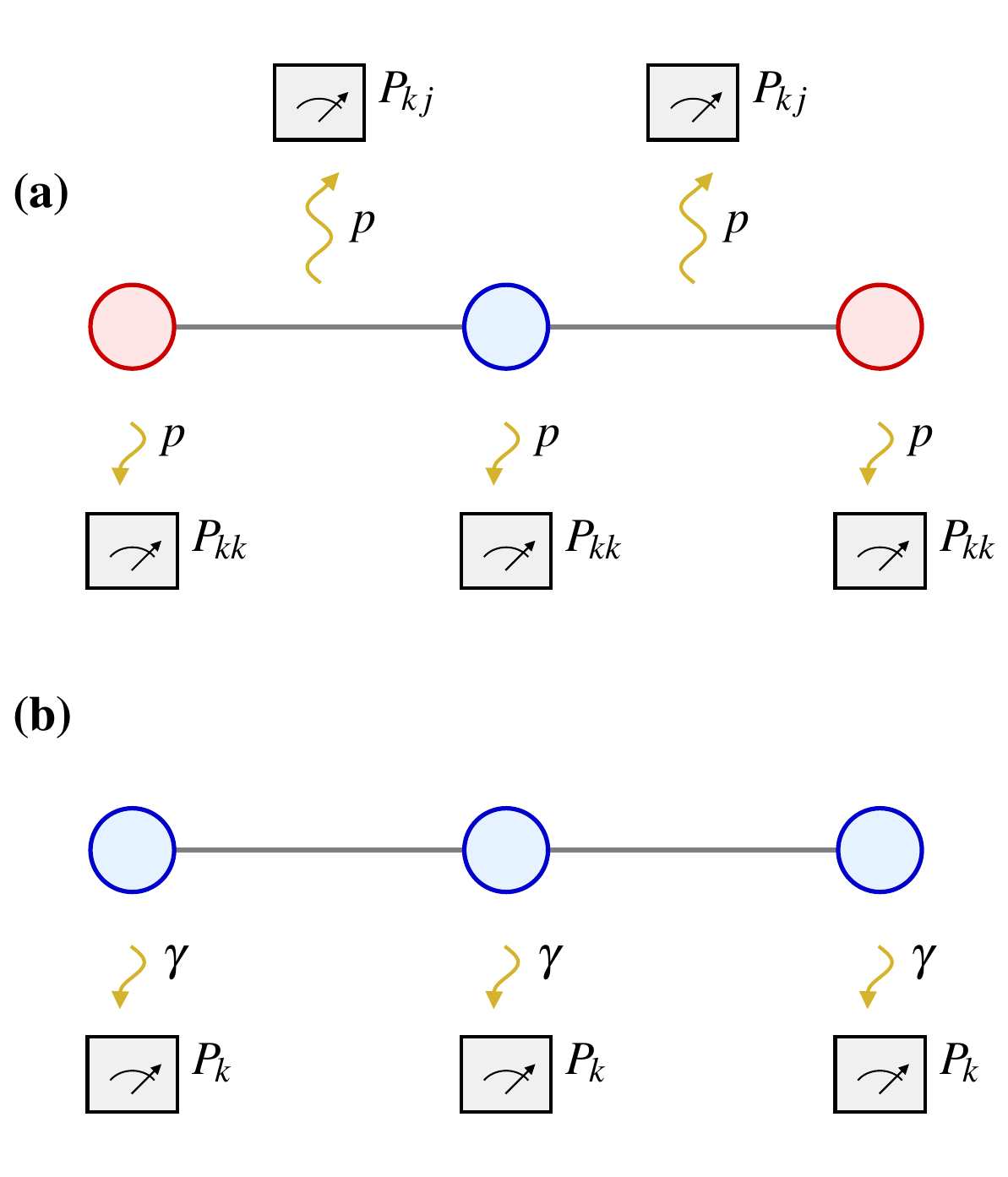}
\caption{Schematic illustration of the postselected of CTQW on a line graph (a) QSW decoherence arises from system environment coupling (strength $p$) through jump operators that cause on-site dephasing ($P_{kk}$) as well as hopping between nodes ($P_{kj}$), both of which are monitored. The graph is color-coded to highlight the postselected steady state: low-degree edge nodes are shown in red (indicating localization), while bulk nodes are blue. (b) Haken-Strobl decoherence (strength $\gamma$) which cause only onsite dephasing ($P_{k}$), which is registered by the detectors. In both cases, the decoherence-induced jumps are depicted by yellow wavy lines.} 
\label{fig:schematic_CTQW}
\end{figure}

QSW model describes a decoherence mechanism where the system-environment coupling induces incoherent transitions between nodes~\cite{QSW}, as illustrated in Fig.~\ref{fig:schematic_CTQW}(a). Beyond its implications for energy and information transport~\cite{mohseni2008environment,dudhe2022testing,govia2017quantum,taketani2018physical}, the QSW class of jump operators has found conceptual applications in advanced theoretical models, including projective simulation models of artificial intelligence, learning agents and quantum inspired decision making~\cite{briegel2012projective,martinez2016quantum}. The jump operators correspond to an edge describing a transition from node $j$ to node $k$ with amplitude $H_{kj}$~\cite{QSW}:

\begin{equation}
P_{kj} = H_{kj}\,|k\rangle \langle j|.
\label{eq:12}
\end{equation}

There is also a degree dependent on-site dephasing due to the jump operators' dependence on the graph's Laplacian. Unlike the pure on-site dephasing model~\cite{haken1973exactly}, this process is explicitly edge (connection) dependent. This introduces a spatial inhomogeneity in the dissipation and becomes crucial for the localization on low-degree nodes that we observe at the heterogeneous networks.

On the other hand, Haken-Strobl decoherence~\cite{haken1973exactly} describes pure dephasing where the environment interacts locally with each node. This model is particularly significant in the study of quantum biology, as it provides a key mechanism for Environment Assisted Quantum Transport, where dephasing can surprisingly enhance the efficiency of energy transfer in photosynthetic complexes~\cite{OptimalEAQT,plenio2008dephasing,caruso2009highly}. The jump operators are projectors onto the local site basis $|k\rangle$~\cite{haken1973exactly,bressanini2022decoherence} : 
\begin{equation}
P_{k} = |k\rangle \langle k|.
\label{HSeqn}
\end{equation}
Physically, this corresponds to the environment continuously monitoring the node (site) of the walker, as illustrated in Fig.~\ref{fig:schematic_CTQW}(b)

\section{Results}
\label{results} 
\begin{figure*}[htbp]
\centering
\includegraphics[width=0.77\textwidth]{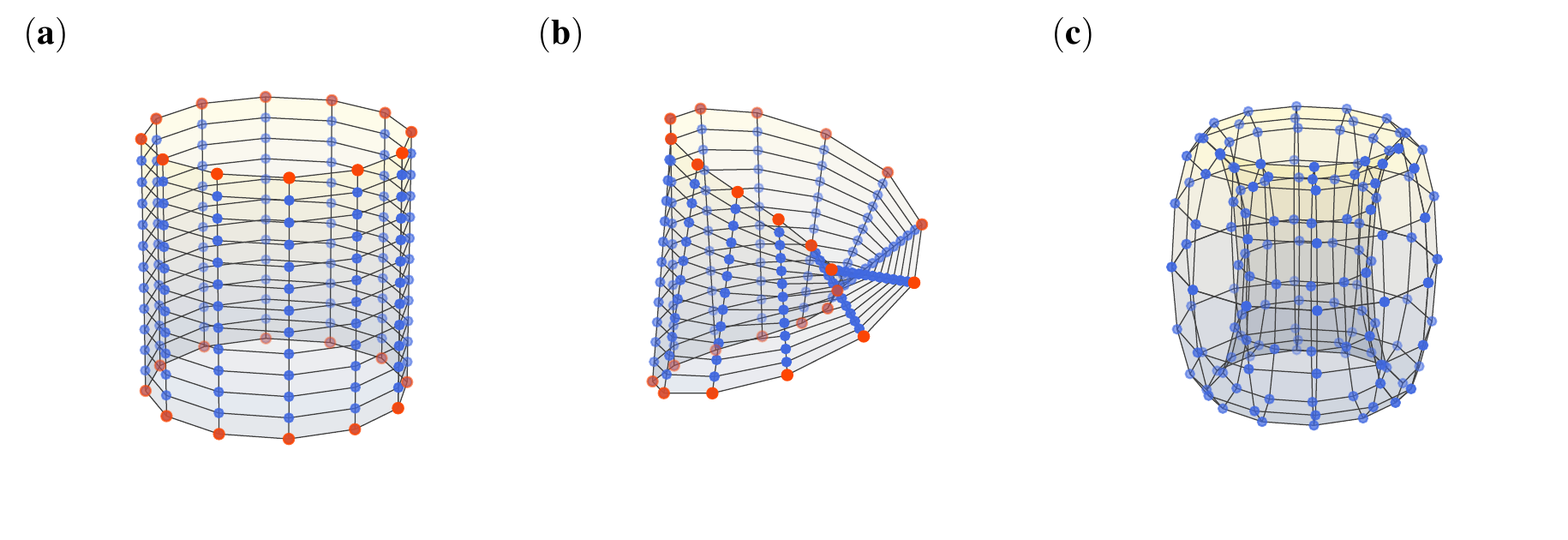}
\caption{Schematic representation of the three network topologies studied. (a) The cylinder topology, having two distinct boundaries composed of degree-3 edge nodes (red). (b) The Möbius strip, which possesses a single continuous boundary of degree-3 edge nodes (red). (c) The torus, a regular graph with no boundaries, where all nodes are part of the bulk and have a uniform degree of 4 (blue). Majority of the nodes shown in blue colors has the same degree 4.}
\label{fig:lattice}
\end{figure*}

Having set-up the formalism, now we proceed to discuss the steady state properties under these decoherence mechanisms. To particularly investigate the interplay between network structure and dynamics, we analyze the system on three representative topologies shown in Fig.~\ref{fig:lattice}: the cylinder, the Möbius strip, and the torus. These topologies are defined on a 2D lattice with a fixed size of N=25 nodes, with the walker initially localized at the center of the graph. Specifically, the cylinder, torus and Möbius strip are constructed by applying periodic boundary conditions along one axis, on both axes, and with an inverted periodic connection, respectively. Both the cylinder and Möbius strip exhibit weak structural heterogeneity due to the presence of boundary (edge) nodes with reduced degree, whereas the torus remains fully homogeneous (regular graph) since all nodes possess identical degrees and an equivalent local connectivity. These graphs were chosen to identify how degree heterogeneity drives localization at low-degree nodes under steady-state conditions. 

\subsection{Topology Dependent Localization from Postselection }
\label{sec:QSW}

First we consider NLME under the framework of QSW where the relevance of network topology becomes prominent in the presence of postselection. QSW is an interpolation between the classical and quantum continuous-time walks which is governed by LME of the form~\cite{QSW}:
\begin{equation}
\begin{split}
\frac{d\rho}{dt}
= -i(1-p)[H,\rho]
+ p\sum_{k,j} \Bigl(
-\tfrac{1}{2}\{P_{kj}^\dagger P_{kj}, \rho\}  \\
\qquad\qquad
+ \,P_{kj}\rho P_{kj}^\dagger
\Bigr).
\end{split}
\label{eq:QSW_LME}
\end{equation}
Here, $p$ is a parameter controlling the relative strength of the decoherence and $P_{kj}$ are jump operators given by Eq.~\eqref{eq:12}. To incorporate postselection into the QSW framework, we apply the quantum trajectory method~\cite{liu2025lindbladian} to the QSW LME in Eq.~\eqref{eq:QSW_LME}, resulting in the corresponding NLME:
\begin{equation}
\begin{split}
\frac{d\rho}{dt}
= -i(1-p)[H,\rho]
+ p\sum_{k,j} \Bigl(
-\tfrac{1}{2}\{P_{kj}^\dagger P_{kj}, \rho\}  \\
\qquad\qquad
+ (1-\eta)\,P_{kj}\rho P_{kj}^\dagger
+ \eta\,\mathrm{Tr}(P_{kj}^\dagger P_{kj}\rho)\,\rho
\Bigr).
\end{split}
\label{eq:15}
\end{equation}

To investigate the dynamics of Eq.~\eqref{eq:15}, we obtain exact numerical solutions on different graph topologies. The numerical integration is performed using fourth-order Runge-Kutta method with an integration step size of $0.01$. To prevent small numerical drift, the trace of the density matrix is renormalized at each time step, and its positivity is continuously verified. The system is evolved until it reaches a steady state, which is numerically identified when the maximum rate of change among all density matrix elements falls below a threshold of $10^{-5}$ for initial states localized at the center of the graph.

While the full analytical solution of the steady state is difficult, we obtain certain consistency conditions on the steady state in Appendix \ref{appendix_QSW}. First we obtain the necessary and sufficient condition for the maximally mixed state $\rho^{ss}=\tfrac{\mathbb{I}}{N}$ to be the system's steady state, in terms of the degrees of the nodes $D_k$ :

\begin{equation}
\eta \sum_{k}\frac{D_{k}^2 + D_{k}}{N}\,|k\rangle\langle k|= \eta \frac{\sum_{j}(D_{j}^2 + D_{j})}{N^{2}}\,\mathbb{I}.
\label{eq:20}
\end{equation}

\begin{figure*}[htbp]
\centering
\includegraphics[width=0.8\textwidth]{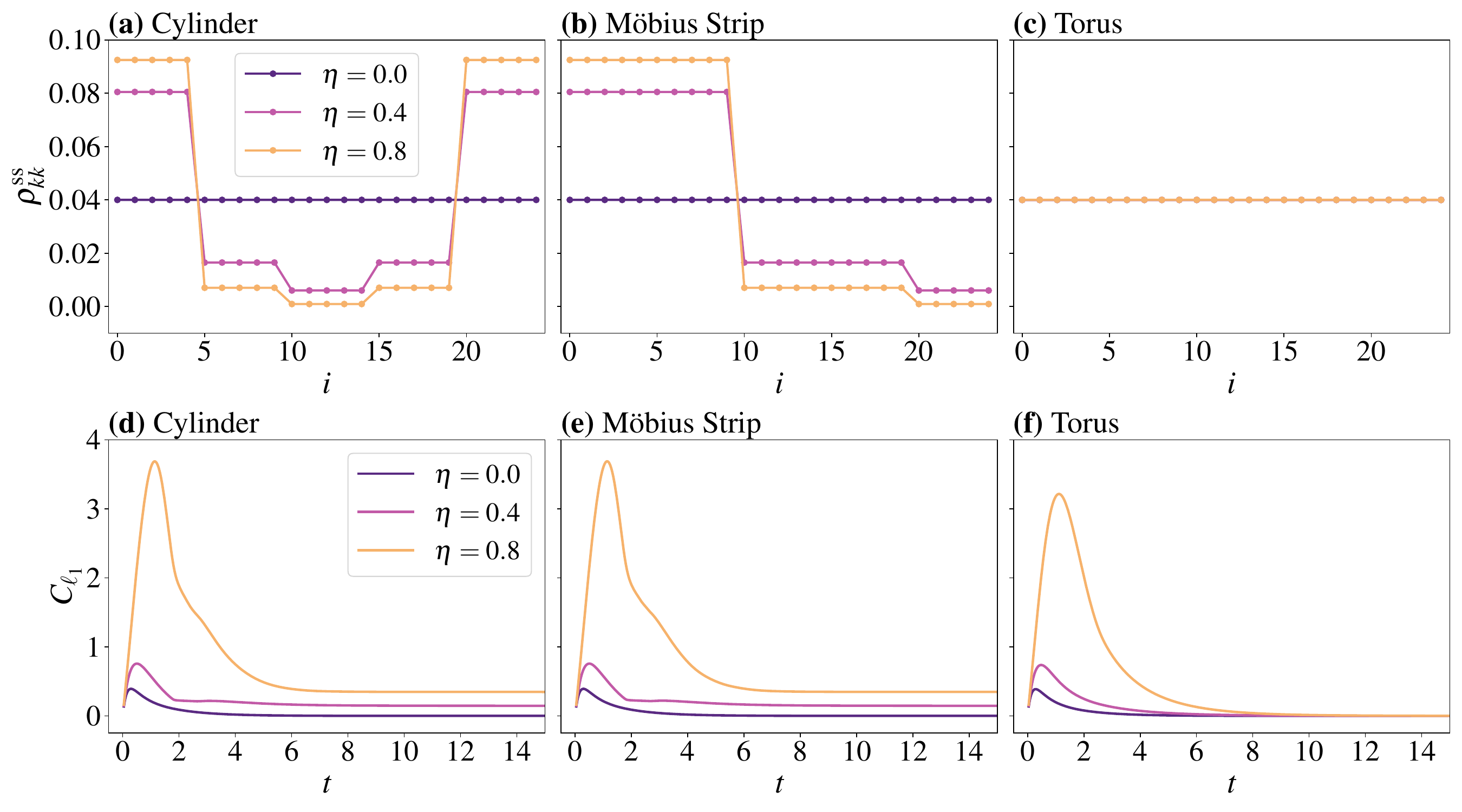}
\caption{Steady-state probability distributions and $\ell$-norm of coherence for CTQW under QSW decoherence. (a) On the Cylinder and (b) the Möbius strip, postselection induces strong localization on the low-degree edge nodes (degree 3). The elevated probability on the first layer of bulk nodes is due to their connection to these highly populated edge nodes. (c) On the edgeless Torus, where all nodes are in the bulk (degree 4), the distribution remains uniform, showing that degree heterogeneity is essential for the localization. In all cases, the initial state was localized at a center node and decoherence strength, $p=0.5$.}
\label{fig:steadystate_QSW}
\end{figure*}

In the absence of postselection $\eta=0$, the equation is trivially satisfied. This confirms that for standard QSW decoherence, the system's long-time steady state is invariably the uniform distribution (Fig.~\ref{fig:steadystate_QSW}(a), (b) and (c)), with coherence decaying to zero (Fig.~\ref{fig:steadystate_QSW}(d), (e) and (f)) and is consistent with previous studies~\cite{stabilityctqw}. 
When postselection is active ($0<\eta<1$), however, Eq.~\eqref{eq:20} only holds if the diagonal operator  $\sum_{k} (D_{k}+D_{k}^2)|k\rangle \langle k|$ is a scalar multiple of identity. This condition is met only if all nodes in the graph have the same degree. Thus for $0<\eta<1$, the uniform distribution of probabilities remain a steady state only for regular graphs. These results precisely validates our numerical findings for the torus topology Fig.~\ref{fig:steadystate_QSW}(c) and (f) which, being regular, maintains a uniform steady state regardless of $\eta$, provided that the detector is imperfect. For heterogeneous topologies like the cylinder and Möbius strip Fig.~\ref{fig:steadystate_QSW}(a) and (b), this condition is violated, where the uniform distribution as steady state is not a solution.

\begin{figure}[htbp]
\centering
\includegraphics[width=0.43\textwidth]{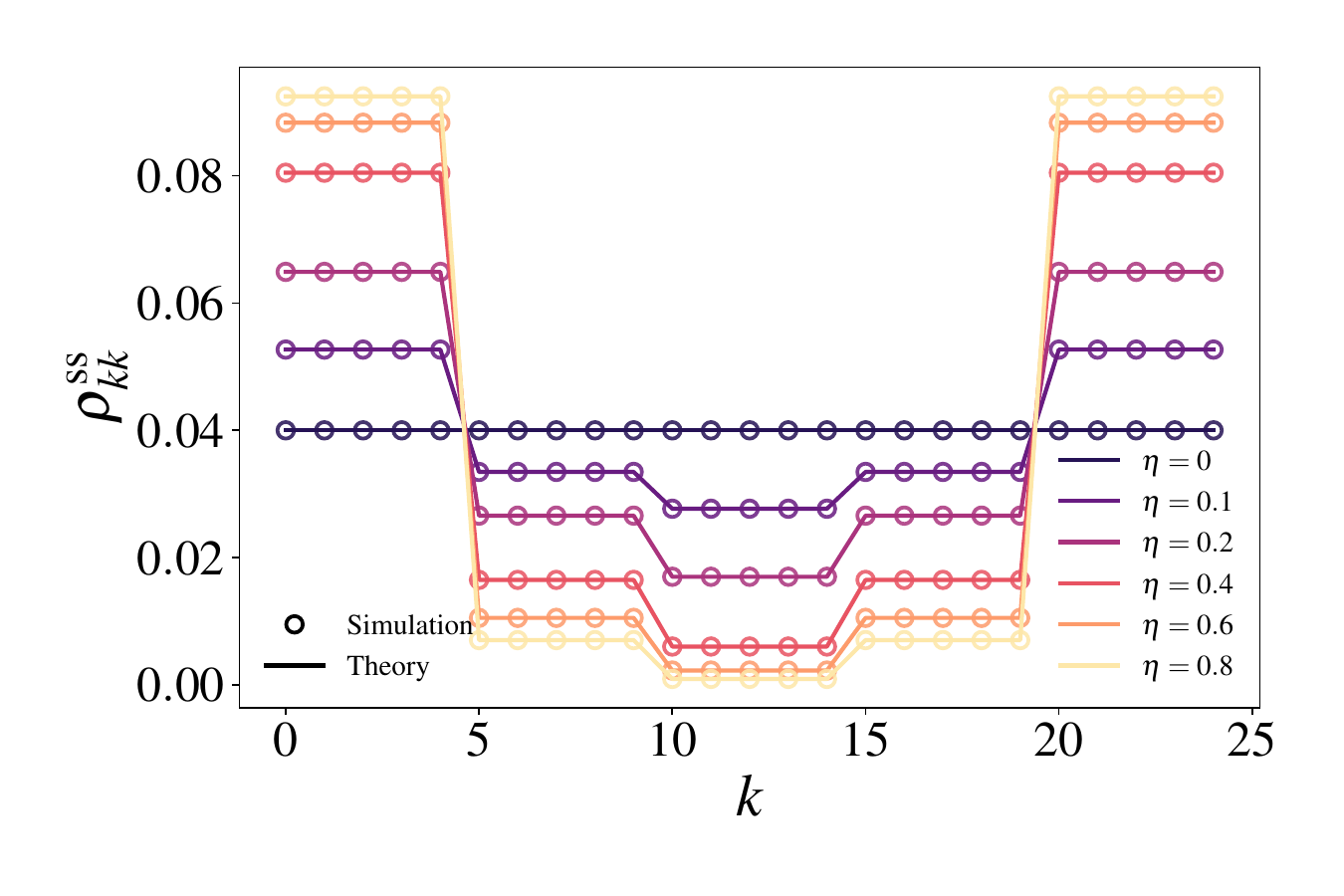}
\caption{Numerical verification of the derived steady-state constraint (Eq.~\eqref{eq:25}). The steady-state probabilities $\rho^{ss}_{kk}$ obtained from time evolution (markers) are compared against the RHS of analytically derived Eq.~\eqref{eq:25} (dashed lines) for the cylinder graph. The plot shows that the simulated steady state strictly obeys the derived analytical condition for all tested values of $\eta$ (p=0.5).}
\label{equation_check}
\end{figure}

In appendix~\ref{appendix_QSW}, we also analyze the diagonal terms of the steady-state condition (Eq.~\ref{eq:19}) and find that the steady state obeys the following constraint for a general heterogeneous graphs:

\begin{equation}
\rho^{\mathrm{ss}}_{kk}=\frac{\sum_{j}A_{kj}\left(p(1-\eta)\rho^{\mathrm{ss}}_{jj}+2(1-p)\mathrm{Im}(\rho^{\mathrm{ss}}_{kj})\right)}{p\left(D_{k}(1+\eta D_{k})-\eta\sum_{j}(D_{j}^2+D_{j})\rho^{\mathrm{ss}}_{jj}\right)}.
    \label{eq:25}
\end{equation}
 To verify this constraint, we obtain exact numerical solutions of the master equation (Eq.~\eqref{eq:15}) on a  cylinder graph. The numerical results (solid lines) exactly match with the theoretical values (markers) obtained from the condition Eq.~\eqref{eq:25}, as presented in Fig.~\ref{equation_check}, with a maximum absolute error of $10^{-7}$. 
 
The analytical form of Eq.~\eqref{eq:25}, satisfies the consistency condition that $\frac{\mathbb{I}}{N}$ is steady state for $\eta=0$ and for regular graphs. Two factors of the Laplacian play a crucial role in determining the properties of the steady state (Eq.~\eqref{eq:25}): $D_k$ and coupling between each node $A_{kj}$. The dependence on node degree highlights how degree heterogeneity serves as an organizing principle for emergence of non-trivial steady states, a phenomenon also observed in classical dynamics~\cite{meena2023emergent}. The decoherence causes jumps that depends on graph's connections (Eq.~\eqref{eq:12}). The physical role of these connections within the NLME (Eq.~\eqref{eq:15}) is most transparent through the total jump intensity from a given node $k$, which is governed by the operator sum identity $\sum_j P_{jk}^\dagger P_{jk}= (D_k^2+D_k)|k\rangle\langle k|$ (Eq.~\eqref{eq:16}). This highlights that high-degree nodes have a larger jump intensity. Therefore, a higher number of connections lead to more dissipative pathways, suppressing the steady-state probability at a particular node. Consequently, low-degree nodes are preferentially populated. Second, $A_{kj}$ introduces a dependence, where the probability at a specific node is influenced by its immediate neighborhood.

 The interplay between node degree and the dependence  on neighboring node coupling (Eq.~\ref{eq:25}) is further observed in our numerical simulations on heterogeneous networks as shown in Fig.~\ref{fig:steadystate_QSW}(a) and (b). In the cylinder and Möbius strip, the edge nodes possess lowest degree, allowing them to have highest steady-state probability. The neighboring nodes of these edge nodes exhibit a slightly higher probability than the nodes in the bulk of the networks. Thus heterogeneous networks favor localized steady states at edge nodes under postselection. As shown in Fig.~\ref{fig:steadystate_QSW}(d) and (e), the system retains significant $\ell_1$-norm of coherence ($C_{\ell_1}(\rho) = \sum_{ i \neq j}|\rho_{ij}|$)\cite{baumgratz2014quantifying,stabilityctqw} for $0<\eta<1$, whereas it decays to zero under standard LME ($\eta=0$). This finding shows that for networks with degree heterogeneity, postselection is a viable mechanism for preserving coherence. As $\eta$ increases, the localization on low-degree nodes becomes more pronounced, and the amount of retained steady-state coherence is likewise enhanced. 

\begin{figure*}[htbp]
\centering
\includegraphics[width=0.9\textwidth]{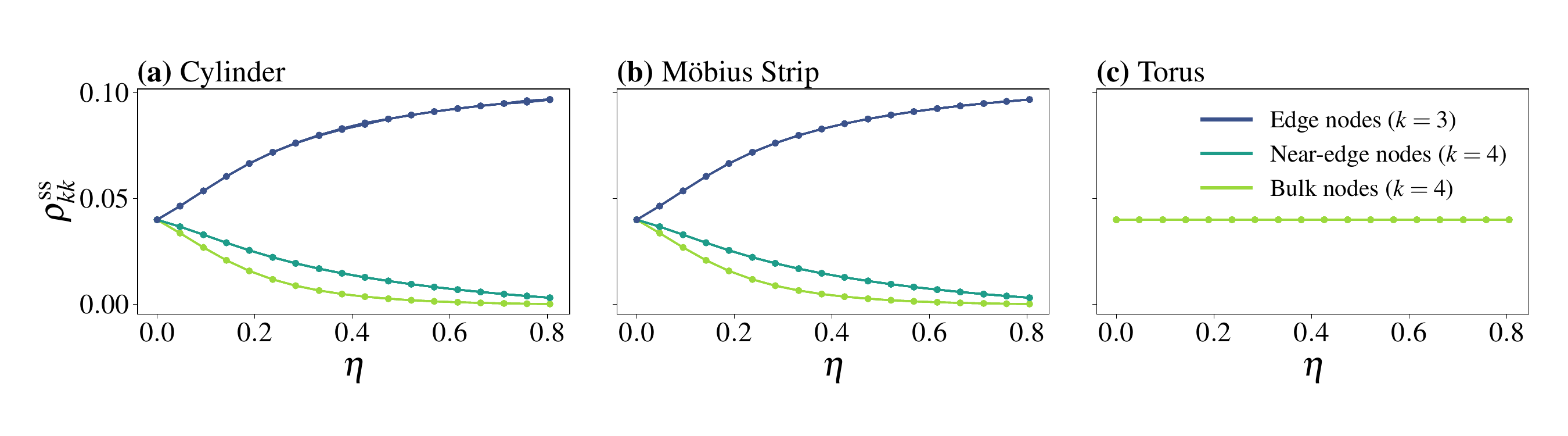}
\includegraphics[width=0.9\textwidth]{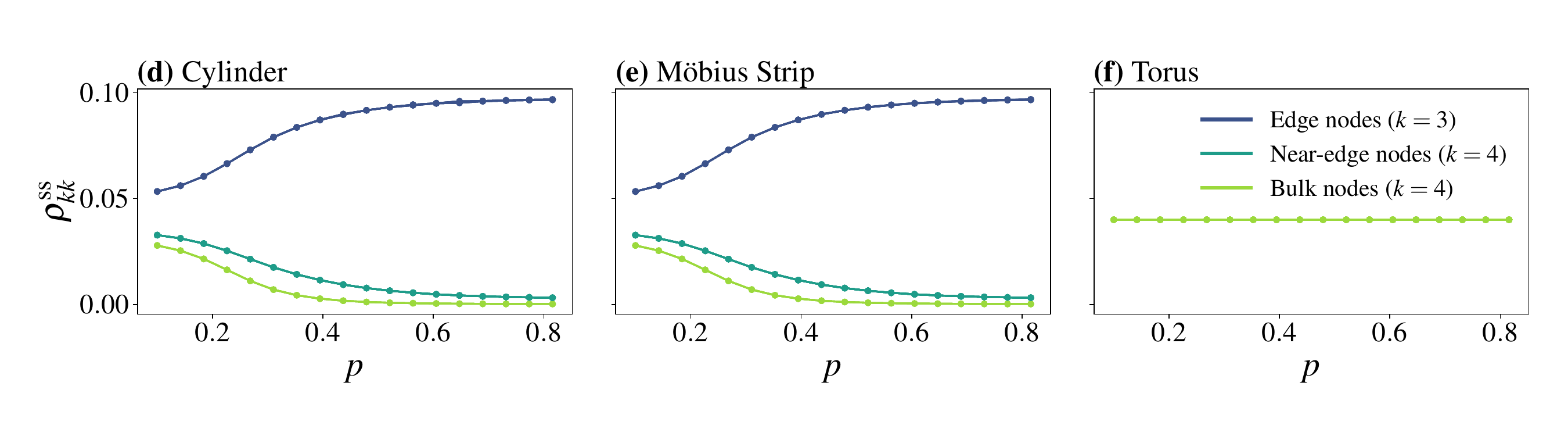}
\caption{ Steady-state probability distributions ($\rho^{ss}_{kk}$) of a CTQW across network nodes under varying decoherence and postselection. The plots illustrate the distribution of steady-state probabilities for a cylinder topology with initial localization is towards one edge. (a)-(c) Steady-state probability distributions for varying efficiency ($\eta$) at fixed strong decoherence ($p = 0.8$)  (d)-(f) Distributions for varying decoherence strength ($p$) at a fixed strong postselection ($\eta=0.8$). }
\label{bifurcation}
\end{figure*}

This low-degree node localized steady state is independent of the initial state. On the cylinder graph, which possesses two distinct edges, initializing the walker near one specific edge yields the same steady-state probability distribution across both edges (Fig.~\ref{bifurcation}(a) and (d)). Because such biased initial state can cause the walker to become transiently trapped in metastable states (Appendix~\ref{initial_state}), a stricter numerical tolerance of $10^{-9}$ is required to ensure convergence to the steady state in these cases. This steady-state localization on the cylinder graph is independent of the initial state, mirroring the steady-state probability distribution of the Möbius strip (Fig.~\ref{bifurcation}(b) and (e)), where the walker concentrates along its single continuous edge. For both of these topologies, the preference for localization at low-degree nodes becomes prominent as either the decoherence parameter $p$ or detection efficiency $\eta$ increases. Conversely, the regular torus (Fig.~\ref{bifurcation}(c) and (f)) maintains a uniform distribution under all parameters. In Appendix~\ref{initial_state}, we numerically verify that the steady state on the cylinder graph is stable and unaffected by the initial localization. 

 To further characterize the localization timescales when the walker on the cylinder graph is initialized at its  center, we detail in Appendix~\ref{transient_dynamics} how the convergence toward the steady state depends on the parameters $\eta$ and $p$. We also show that the localization time scales linearly with system size $N$, causing the survival probability to decay exponentially. Consequently, the accepted fraction of trajectories can be small even for finite networks. Our results therefore delineate the small network regime (as given by the bounds in Eq.~\eqref{eq.C7}) in which the localization phenomenon is experimentally accessible and provide a practical criterion for optimizing the postselection strength and observation time. Moreover, localization on low-degree nodes reduces the instantaneous rejection rate relative to a delocalized state, partially mitigating the sampling cost. The predicted effect is thus most naturally viewed as a heralded phenomenon in small networks, whose observability can be systematically controlled by balancing localization fidelity against acceptance probability.
 
\begin{figure*}[htbp]
\centering
\includegraphics[width=1\textwidth]{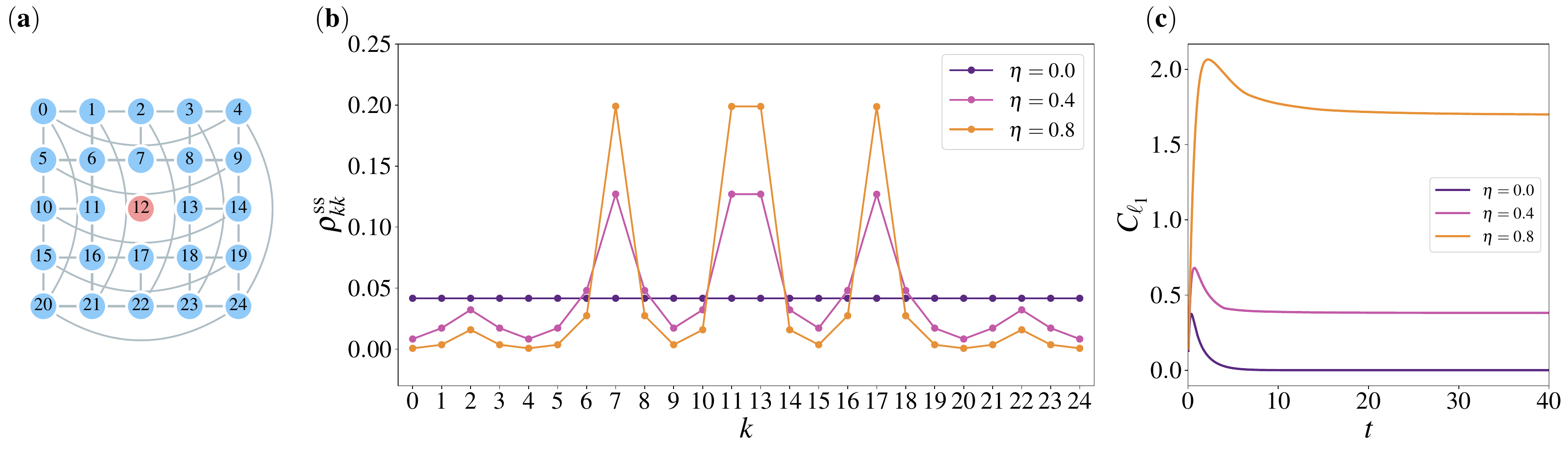}
\caption{ Steady state of a defected torus (single node removal) with decoherence strength $p=0.5$ and initial localization on node index 7. (a) Schematic of the graph. The removed node (index 12) is shown in red. The four immediate neighbors (indices 7, 11, 13, and 17) effectively become low-degree edge nodes due to the removal. (b) Steady-state probability distribution for the defected torus. (c) Time evolution of the $\ell_1$-norm of coherence for varying postselection efficiencies $\eta$.}
\label{torous_defect}
\end{figure*}

To provide definitive numerical proof that this localization is driven by the node degree, rather than geometric boundary effects, we investigated the steady state on a torus with defect (Fig.~\ref{torous_defect}(a)). Removing a single node breaks the regular topology; while the global structure remains toroidal. The four immediate neighbors of the vacancy effectively become low-degree edge nodes. This heterogeneity causes the steady state to transition from the uniform distribution observed in the perfect torus (Fig.~\ref{fig:steadystate_QSW}(c)) to a localized distribution concentrated on these low-degree neighbors (Fig.~\ref{torous_defect}(b)). For moderate and strong postselection ($\eta=0.4$ and $0.8$), the steady-state probability is distributed equally among all four low-degree neighbors, reflecting the local topological symmetry of the defect.  Regarding the coherence dynamics, Fig.~\ref{torous_defect}(c) shows the time evolution of the $\ell_{1}$-norm of coherence. As the $\eta$ increases, the system retains a significantly higher quantum coherence, with decay being strongly suppressed during the transient phase toward the steady state. 

Now we turn to the case of the NLME under Haken-Strobl decoherence. Assuming site independent monitoring efficiency $\eta$,  we substitute the site projection jump operators (Eq.~\ref{HSeqn}) into Eq.~\eqref{eq:1} which significantly simplifies the dynamics. Setting the steady-state condition $\dot{\rho^{ss}}=0$ yields the balance equation:  

\begin{equation}
 i[H,\rho^{ss}]=
- \gamma (1-\eta)(\rho^{ss} - \mathcal{P}(\rho^{ss})),
\label{eq:7a}
\end{equation}

where  $\mathcal{P}(\rho^{ss})= \sum_{k} \rho^{ss}_{kk}|k\rangle \langle k|$ is the projection onto the diagonal elements in the site basis, as detailed in Appendix~\ref{sec:appendix_a}. To determine the form of $\rho^{ss}$, we employ a Hilbert-Schmidt inner product argument to Eq.~\eqref{eq:7a} to show that all the off-diagonal elements vanish. Consequently, for any connected graph and $\eta<1$, the system invariably driven to the unique maximally mixed state, $\rho^{ss}=\frac{\mathbb{I}}{N}$.

\begin{figure}[htbp]
\centering
\includegraphics[width=0.37\textwidth]{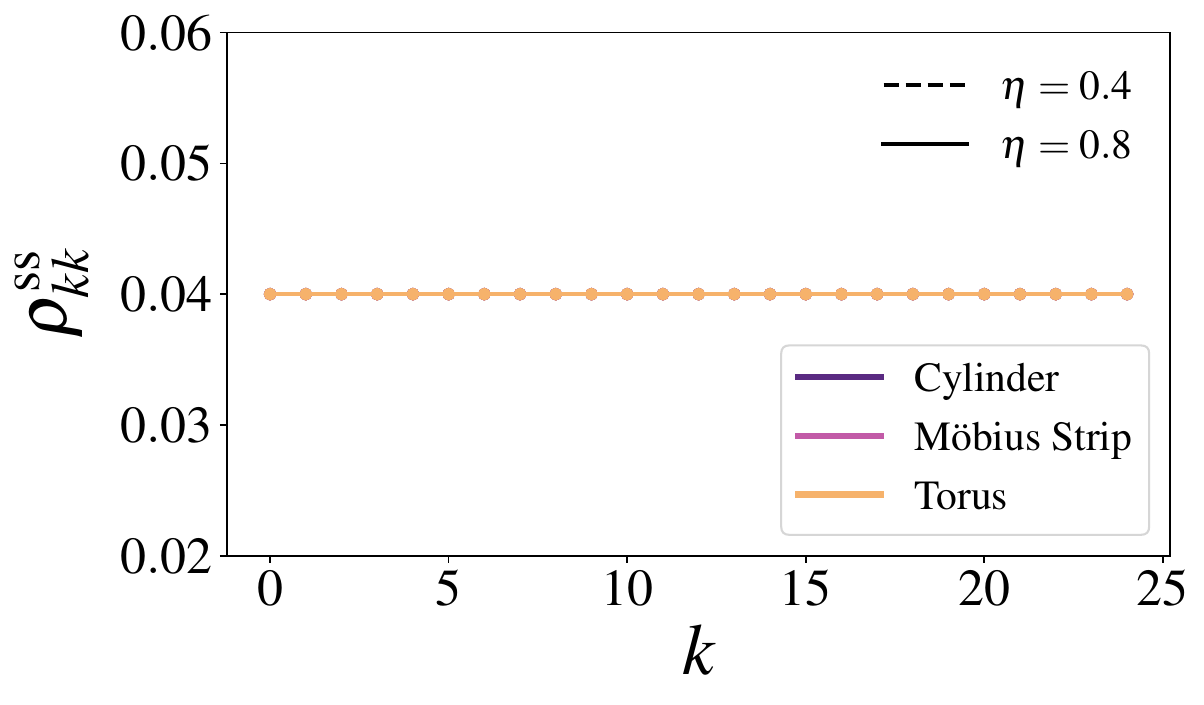}
\caption{Uniform steady-state distributions under Haken-Strobl decoherence. The steady-state probability distributions are shown for Cylinder, Torus, and Möbius strip at postselection efficiencies $\eta=0.4$ and $\eta=0.8$. The final state remains uniform  regardless of the network topologies and detection efficiency $\eta<1$.}
\label{fig:steadystate_HS}
\end{figure}

Our numerical simulation in Fig.~\ref{fig:steadystate_HS} confirm this analytical prediction, showing that the system converges to a uniform probability distribution across all nodes for the cylinder, Möbius strip, and torus topologies, for $\eta<1$. For the specific case where $\eta=0$, which corresponds to the standard Lindblad Master Equation, this result is in full agreement with previously established work~\cite{stabilityctqw}.

This completes our discussion of the steady state of the NLME induced CTQW on heterogeneous networks with single particle states. Next we consider CTQWs on spin networks.

\subsection{Topology Dependent Localization and Entanglement Preservation in Spin Networks}
\label{spinnetwork}

\begin{figure}[htbp]
\centering
\includegraphics[width=0.4\textwidth]{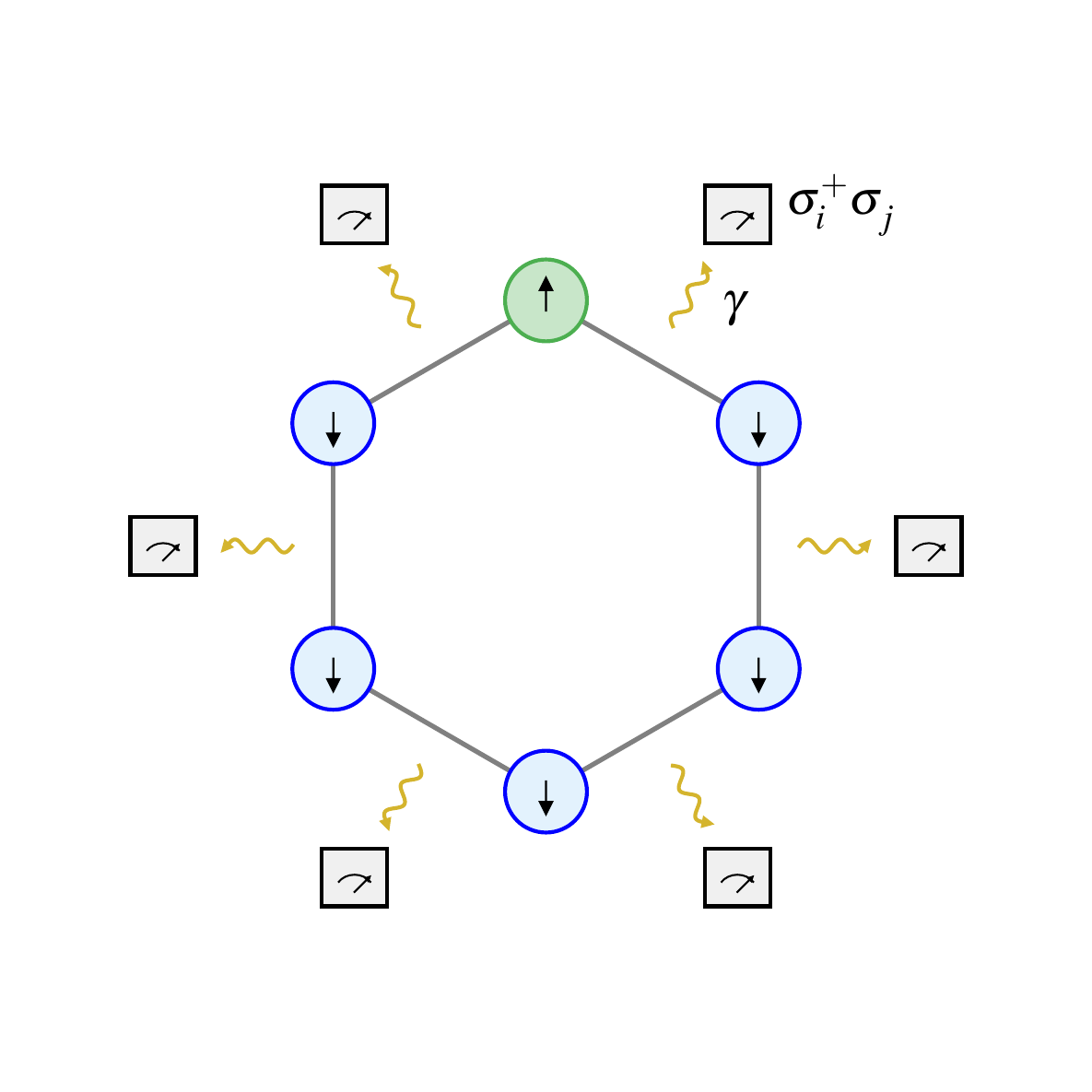}
\caption{Schematic illustration of the postselection experiment in a cycle spin network. A single spin-up excitation (green node) propagates through a background of spin-down sites (blue nodes). Detectors are coupled to the graph edges, continuously monitoring the system to register quantum jumps ($\sigma_i^+\sigma_j^-$) induced by decoherence (strength $\gamma$).}
\label{fig:spin_schematic}
\end{figure}
To investigate whether the localization phenomena persist in the presence of quantum correlations, we analyze the transport of a single spin excitation (spin up) in a network of interacting spins as illustrated in Fig.~\ref{fig:spin_schematic}. The coherent transport of the excitation is governed by the isotropic XY spin Hamiltonian, a foundational model for quantum state transfer in spin chains ~\cite{lieb1961two}:
\begin{equation}
    H = J \sum_{\langle i, j \rangle} \left( \sigma^+_i \sigma^-_j + \sigma^-_i \sigma^+_j \right),
    \label{eq:26}
\end{equation}
where J is the coherent hopping strength between connected spins $⟨i,j⟩$, which is set to unity, and $\sigma^{\pm}$ are the standard spin raising and lowering operators. This Hamiltonian describes the exchange interaction between coupled spins and is the standard model for quantum state transfer protocols~\cite{bose2007quantum,lyakhov2005quantum,lyakhov2006use} and even excitation transport in light-harvesting complexes~\cite{caruso2009highly}. In all these contexts, the transport of a single excitation is the primary figure of merit. We analyze the system's evolution under a decoherence model analogous to the QSW, governed by the Lindblad operator $L_{ij}=\sigma_i^+\sigma_j^-$. This operator describes the environment mediated hopping of an excitation between connected spins~\cite{taketani2018physical}, destroying a spin-up excitation at the source site $j$ and creating one at the target site $i$. By substituting this operator and its adjoint, $L_{ij}^\dagger=\sigma_j^+\sigma_i^-$, into the nonlinear framework (Eq.~\eqref{eq:1}), the resulting NLME is:
\begin{equation}
\begin{split}
\frac{d\rho}{dt} = -i[H, \rho] 
&+ \sum_{\langle i,j \rangle} \gamma \Big(
    - \frac{1}{2} \{ L_{ij}^\dagger L_{ij}, \rho \} + (1-\eta) \, L_{ij}\rho L_{ij}^\dagger \\
&\quad + \eta \, \mathrm{Tr}(L_{ij}^\dagger L_{ij}\rho) \rho
\Big).
\end{split}
\label{eq.qubit_qsw}
\end{equation}

\begin{figure*}[htbp]
\centering
\includegraphics[width=0.87\textwidth]{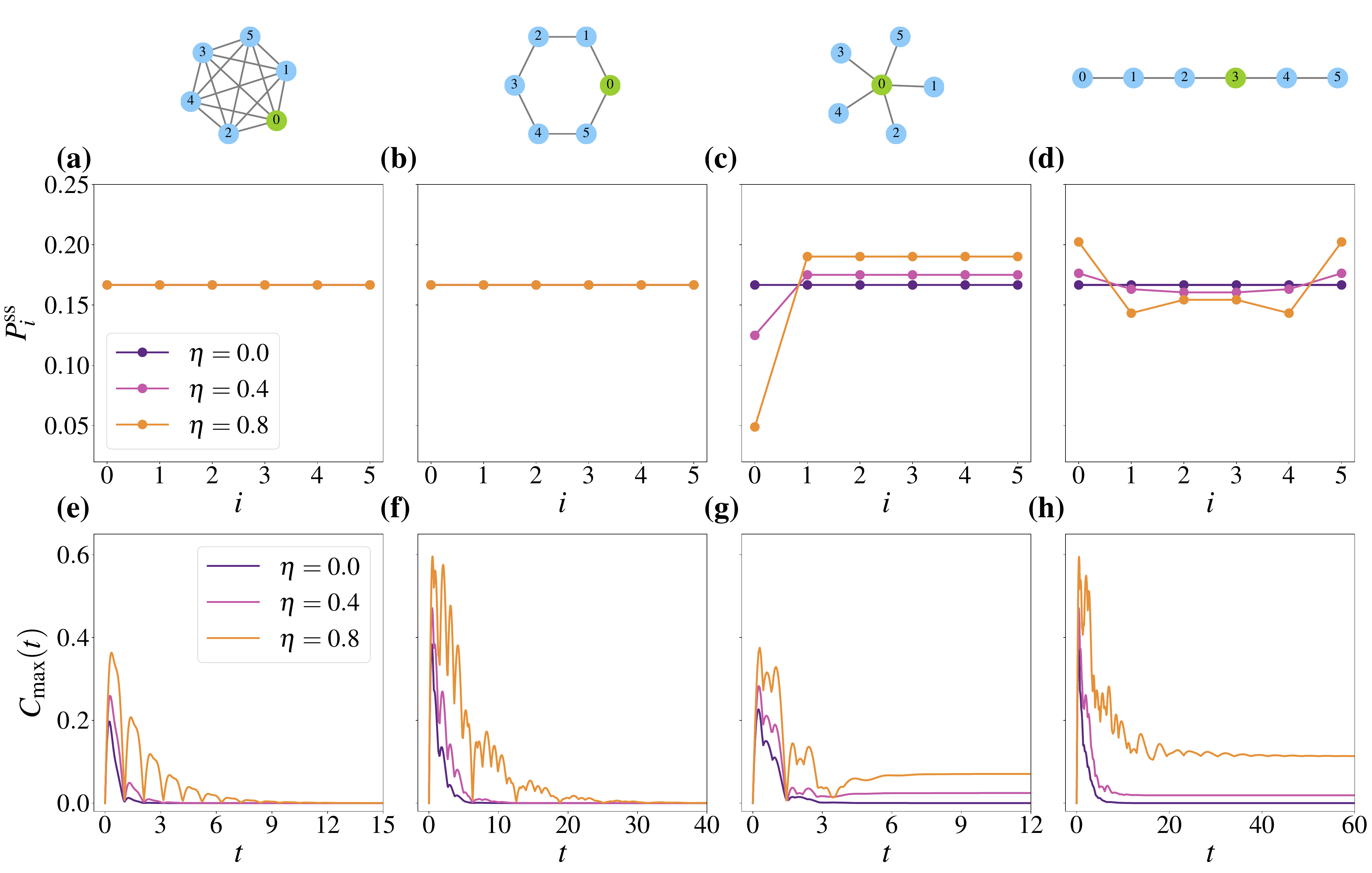}
\caption{ Steady-state spin excitation probability ($P^{ss}_{i}$) at node $i$ and maximum concurrence ($C_{max}(t)$) in spin networks. We investigate spin excitation transport on four network topologies: two with heterogeneous degrees (Line and Star) and two with homogeneous degrees (Complete and Cycle). In these topologies, the nodes with initial spin-up excitation are highlighted in green. For the Star graph, the central hub (index 0) has degree 5, while all peripheral nodes have degree 1. In the Line graph, the endpoints (indices 0 and 5) have degree 1, while bulk nodes have degree 2. (a)-(d) Steady-state probability distributions under QSW decoherence for varying detection efficiency $\eta$. Localization emerges on low-degree nodes for heterogeneous graphs (Star, Line) but remains uniform for regular graphs (Complete, Cycle). (e)-(f) Time evolution of the maximum pairwise concurrence for different efficiencies $\eta$, showing the persistence of entanglement in the steady states localized on edge nodes; different time scales are chosen to capture the transient dynamics.}
\label{fig:steadystate_qubit}
\end{figure*}

To elucidate the interplay between network topology and the nonlinear dynamics of the NLME, we simulate the evolution of 6-spin coupled dynamics across the four distinct topologies shown in Fig.~\ref{fig:steadystate_qubit}. These include the complete and cycle networks, which consists of nodes with uniform degrees (homogeneous), and the star and line networks, which have nodes with varying degrees (heterogeneous). By utilizing these small networks, we can transparently analyze the postselected dynamics of the open spin system and track the pairwise entanglement preservation. Our numerical results reveal topology dependent distinction in the steady state under QSW like decoherence (Sec.~\ref{sec:QSW}). As shown in Fig.~\ref{fig:steadystate_qubit}(a)-(b), for regular networks (Complete and Cycle), the presence of postselection ($0<\eta<1$) does not affect the system's steady state and so relaxes it to a uniform distribution. In contrast, heterogeneous networks (Fig.~\ref{fig:steadystate_qubit}(c)-(d)) exhibit edge localization in this regime. The steady state probability increases on low-degree nodes monotonically with the post selection efficiency $\eta$. 

A key question is whether the steady state characterized by postselected localization at low-degree nodes preserves or suppresses quantum entanglement. While single particle coherence measures, such as the $\ell_1$-norm of coherence, suffice to quantify the superposition of a single walker, they do not help to quantify  non-local quantum correlations (entanglement) in an open spin system. To quantify this, we compute the pairwise concurrence $C_{ij}$ between spins $i$ and $j$. Since our system is in a mixed state $\rho(t)$ due to the decoherence, we employ the standard Wootters formula~\cite{wootters1998entanglement}. Recently, the experimental quantification of concurrence in solid-state systems has also become feasible, as demonstrated in the quasi one-dimensional spin-1/2 chain $Cs_2CoCl_4$~\cite{laurell2021quantifying}. For any pair of spins at position $i$ and $j$, we first construct the two-site reduced density matrix $\rho_{ij}$ 
by tracing out all other sites. The concurrence is then calculated as: $C_{ij} = \max\left(0, \lambda_1 - \lambda_2 - \lambda_3 - \lambda_4 \right)$, where $\lambda_k$ are the square roots of the eigenvalues of the matrix $R = \rho_{ij} \tilde{\rho}_{ij}$
in descending order, where $\tilde{\rho}_{ij} = (\sigma_i^y \otimes \sigma_j^y)\, \rho_{ij}^*\, (\sigma_i^y \otimes \sigma_j^y)$ and $\sigma_i^y$ is the standard Pauli matrix acting on $i^{th}$ spin. Figs.~\ref{fig:steadystate_qubit}(e)-(h) present the maximum pairwise concurrence. For regular networks (Fig.~\ref{fig:steadystate_qubit}(e)-(f)), the entanglement decays to zero as the system relaxes to the maximally mixed state. However, on heterogeneous networks (Fig.~\ref{fig:steadystate_qubit}(g)-(h)), entanglement is not completely destroyed, which means 
the postselection effectively preserves entanglement. Notably, the line graph preserves more entanglement than the star network.  Since the star network is an extremely heterogeneous network, its high-degree central node induces more dissipation. The dependence of entanglement dynamics on node degree is further evidenced by transient dynamics across all graphs (Fig.~\ref{fig:steadystate_qubit}(e)-(h)). Specifically networks with high-degree nodes, such as complete and star graphs, attain a significantly lower peak of maximum concurrence than low-degree networks, such as cycle and line graphs. Crucially, our results suggest that for single-spin excitations in networks, degree heterogeneity sustains finite steady-state pairwise entanglement under postselection.

\section{Conclusion}
\label{conclusion}

In this study, we show that the interplay between nonlinear decoherence mechanisms and network structure significantly influences the dynamics of CTQWs. In particular, the emergence of localization under postselection strongly depends on the type of decoherence. Under QSW decoherence with imperfect detection, postselection breaks the dynamical balance on networks with heterogeneous degree distributions, resulting in robust localization at low-degree nodes. In contrast, for symmetric on-site dephasing described by the Haken–Strobl model, the nonlinear contributions cancel out, and the system relaxes to a uniform steady state, provided the detector is not perfect.

Crucially, we show that the effective non-reciprocity required for this localization does not need to be explicitly built into the Hamiltonian. Instead, it emerges dynamically from the interplay of a standard hopping Hamiltonian, the specific environmental coupling (QSW), and the measurement postselection process. Expanding this analysis to the transport of a single spin excitation in a network of interacting spins, we confirmed that localization on low-degree nodes emerges from hopping processes even in the presence of quantum entanglement. Importantly, we find that in the localized regime, quantum coherence is preserved; in the context of spin excitation transport, this manifests as the preservation of pairwise entanglement between spins in the network.

This decoherence-dependent localization offers a mechanism for quantum state engineering and control. The ability to steer a quantum walker to the boundaries of a network by tuning the postselection provides a powerful tool. This capability could be used to control the output state of a quantum search algorithm~\cite{xing2024quantum} or to guide quantum information to specific nodes within a quantum processor~\cite{bose2007quantum}. By designing networks with specific degree distributions, one can effectively pre-program the final location of the quantum walker. Ultimately, our work demonstrates that postselection is not merely a passive filtering process but an active tool for manipulating quantum transport, providing a physically grounded platform for exploring and controlling the rich phenomena of non-Hermitian physics. Looking forward, it would be particularly insightful to explore the interplay between postselection-induced nonlinearities and the underlying topology of the Liouvillian spectrum~\cite{okuma2023non,ding2022non,banerjee2023non}. Such a study would bridge our findings with an active field of research in condensed matter physics where networks are treated as lattices.

\begin{acknowledgments}
CM acknowledges support from the Anusandhan National Research Foundation (ANRF) India (Grants Numbers SRG/2023/001846 and EEQ/2023/001080) and Department of Science and Technology, India (INSPIRE Faculty Grant Number IFA19-PH248). SH would like to acknowledge funding support from the ANRF Prime Minister Early Career Grant(PM-ECRG): ANRF/ECRG/2024/002083/PMS.
\end{acknowledgments}
\section*{Data availability}
The data are generated through numerical simulations of the dynamical equations presented in this work. The codes supporting the findings of this study are available from the authors upon reasonable request. 
\appendix
\section{Steady State Under Quantum Stochastic Walk Decoherence}
\label{appendix_QSW}
To derive the steady state for the QSW decoherence, we simplify Eq.~\eqref{eq:15}. The anti-commutator term  which simplifies while substituting for $k=j$, $|H_{jj}|^2=D_{j}^2$ and $k\neq j$ $\sum_{k}|H_{kj}|^2= D_{j}$ from Eq.~\eqref{eq:13}:
\begin{align}
    \sum_{k,j}\{P_{kj}^\dagger P_{kj}, \rho\}&=\sum_{k,j}\{|H_{kj}|^2|j\rangle \langle k|k\rangle \langle j|,\rho\}\nonumber\\&=\{\sum_{j}\left( \sum_{k}|H_{kj}|^2 \right)|j\rangle \langle j|,\rho\}\nonumber\\&=\{\sum_{j}(D_{j}^2+D_{j})|j\rangle \langle j|,\rho\}.
    \label{eq:16}
\end{align}
Then, the jump term can be simplified by substituting ~\eqref{eq:13}:
\begin{align}
\sum_{k,j}(1-\eta)P_{kj}\rho P_{kj}^\dagger 
&= (1-\eta)\sum_{k,j}|H_{kj}|^2 |k\rangle \langle j| \rho |j\rangle \langle k|\nonumber \\
&= \sum_{k,j} (1-\eta) |H_{kj}|^2 \rho_{jj} |k\rangle \langle k|\nonumber \\
&= (1-\eta)\sum_{k} \left( D_{k}^2 \rho_{kk} + \sum_{j} A_{kj} \rho_{jj} \right) |k\rangle \langle k|.
\label{eq:17}
\end{align}

To find the trace term, substitute $H_{kj}$ values from ~\eqref{eq:13}.
\begin{align}
    \sum_{k,j} \eta\, \mathrm{Tr}(P_{kj}^\dagger P_{kj}\rho)\,\rho &= \eta\, \mathrm{Tr}\left(\sum_{j}(D_{j}^2+D_{j})|j\rangle \langle j|\rho\right)\rho \nonumber\\&=\eta\left(\sum_{j}(D_{j}^2+D_{j})\rho_{jj}\right)\rho
    \label{eq:18}.
\end{align}
The steady state equation becomes by imposing the condition $\dot{\rho}^{ss}= 0$ and substituting Eq.~\eqref{eq:16}, Eq.~\eqref{eq:17} and Eq.~\eqref{eq:18}  in Eq.~\eqref{eq:15} :

\begin{equation}
\begin{split}
i(1-p)[H,\rho^{ss}] 
&= -\tfrac{p}{2}\Bigl\{\sum_{k}(D_{k}^2 + D_{k})|k\rangle\langle k|,\rho^{ss}\Bigr\} \\
&\quad +\, p(1-\eta)\sum_{k} \biggl( D_{k}^2 \rho^{ss}_{kk} \\
&\qquad\quad + \sum_{j} A_{kj}\rho^{ss}_{jj} \biggr)|k\rangle\langle k| \\
&\quad +\, p\eta\Biggl(\sum_{j}(D_j^2 + D_j)\rho^{ss}_{jj}\Biggr)\rho^{ss}.
\end{split}
\label{eq:19}
\end{equation}

To analyze the nature of steady state probabilities, $\rho^{ss}_{kk}$, let's analyze the diagonal elements of the full steady-state equation, Eq.~\eqref{eq:19}. Since $H$ is symmetric and $\rho$ is hermitian, the commutator on right hand side of Eq.~\eqref{eq:19} becomes,
\begin{equation}
    [H,\rho_{ss}]_{kk}= -2i\sum_{j} H_{kj}Im(\rho^{ss}_{kj}).
    \label{eq:21}
\end{equation} The anti-commutator term becomes in Eq.~\eqref{eq:19}:
\begin{equation}
    \{\sum_{k}(D_{k}^2+D_{k})|k\rangle \langle k|,\rho^{ss}\}_{kk}= 2(D_{k}^2+D_{k}) \rho^{ss}_{kk}.
    \label{eq:22}
\end{equation}
Substituting Eq.~\eqref{eq:13} into Eq.~\eqref{eq:21}, and subsequently inserting this result along with Eq.~\eqref{eq:22} into Eq.~\eqref{eq:19}, yields:
\begin{equation}
\begin{split}
2(1-p)\sum_{j} A_{kj}\,\mathrm{Im}(\rho^{ss}_{kj})
=\, & p(D_{k}^2 + D_{k})\,\rho^{ss}_{kk}  \\
& - p(1-\eta)\!(
D_{k}^2\,\rho^{ss}_{kk}
+ \sum_{j} A_{kj}\rho^{ss}_{jj}
) \\
& - p\eta\!\left(
\sum_{j}(D_{j}^2 + D_{j})\rho^{ss}_{jj}
\right)\rho^{ss}_{kk}.
\end{split}
\label{eq:23}
\end{equation}  Rearranging Eq.~\eqref{eq:23} gives the steady-state probability as:
\begin{equation*}
\rho^{ss}_{kk}=\tfrac{\sum_{j}A_{kj}\left(p(1-\eta)\rho^{ss}_{jj}+2(1-p)Im(\rho^{ss}_{kj})\right)}{p\left(D_{k}(1+\eta D_{k})-\eta\sum_{j}(D_{j}^2+D_{j})\rho^{ss}_{jj}\right)}.
\end{equation*}

\section{Dependence of the Steady State on Initial Conditions}
\label{initial_state}

\begin{figure}[H]
\centering
\includegraphics[width=0.35\textwidth]{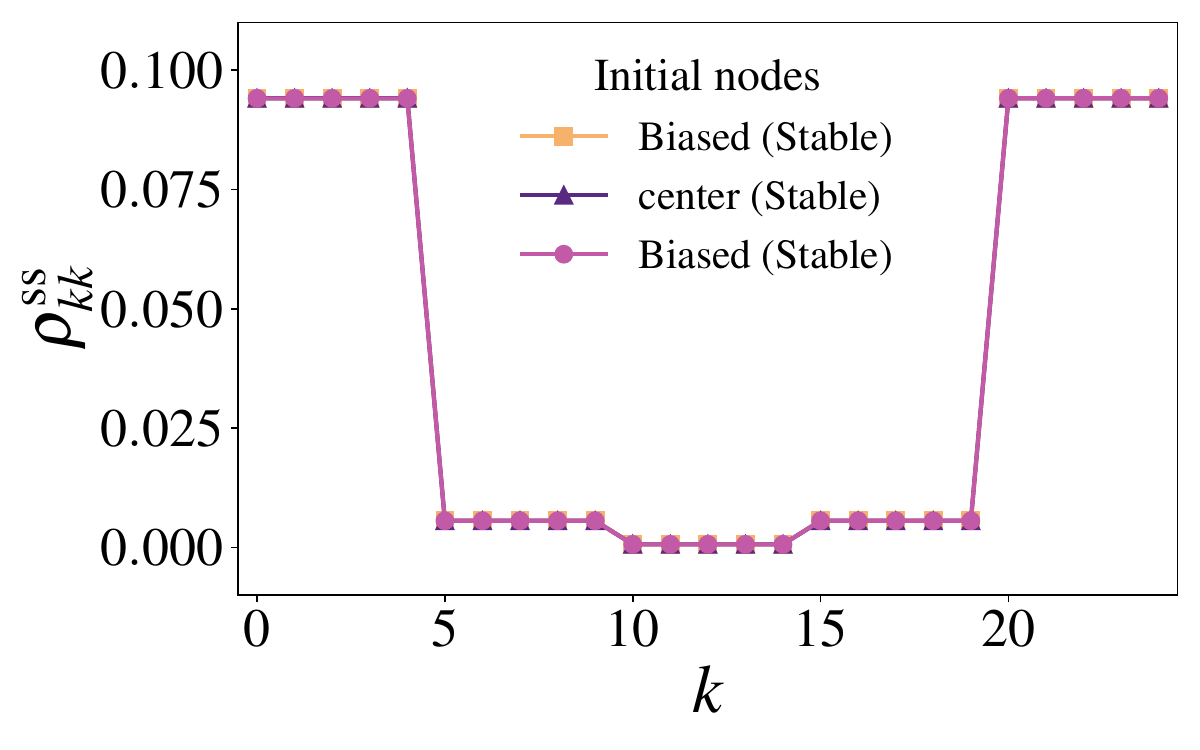}
\caption{Steady-state probabilities for different initial states. For the cylinder graph, even under strong decoherence and efficiency ($p=0.7$ and $\eta=0.7$), the steady-state distribution is identical and dynamically stable regardless of the initial conditions.}
\label{stability plot}
\end{figure}
To confirm that the steady state is independent of the initial state, even when biased to one edge of the cylinder graph, and that the resulting steady state is stable, we employ the linear stability analysis. We first compute the steady state $\rho^{ss}$ by numerically integrating Eq.~\eqref{eq:15}, $\dot{\rho}=\mathcal{L(\rho)}$, until the condition $\mathcal{L}(\rho^{ss})=0$ is satisfied within a stricter numerical tolerance of $10^{-9}$. We then introduce a small perturbation to the steady state, as $\rho=\rho^{ss}+ \sum_b k_b\lambda_b$, where $k_b$ are real coefficients and $\lambda_b$ are the Hermitian, traceless generalized Gell-Mann matrices~\cite{levi2020instability}. The elements of the corresponding Jacobian matrix are given by~\cite{buks2021stability}:
$J_{ab}=\frac{1}{2}Tr(\frac{\partial \mathcal{L}}{\partial k_b}\lambda_a)|_{\rho^{ss}}$. The stability of the steady state is determined by the eigenvalues of this Jacobian matrix. Specifically, if the real parts of all eigenvalues are negative, the steady state is stable. We evaluate the matrix elements numerically using a central finite difference method (step size of $10^{-6}$). As shown in Fig.~\ref{stability plot}, all initial conditions converge to the exact same steady state, which remains dynamically stable even under strong decoherence and detection efficiency.

\begin{figure}[H]
\centering
\includegraphics[width=0.5\textwidth]{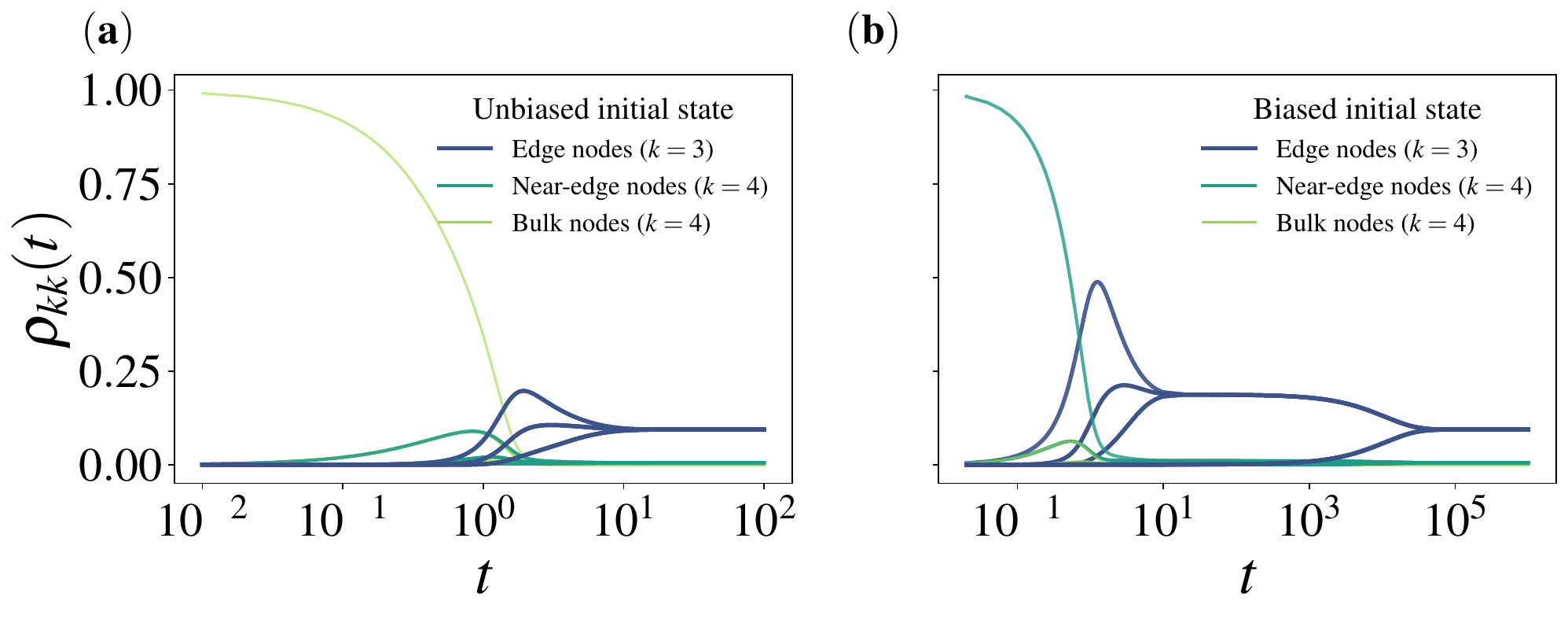}
\caption{Transient dynamics for different initial states under strong decoherence ($p=0.7$) and detection efficiency ($\eta=0.7$). (a) Time evolution for a centrally localized (unbiased) initial state, showing rapid convergence to steady state. (b) Time evolution for an initial state biased toward one edge, showing a delay in reaching the steady state.}
\label{transient plot}
\end{figure}

In Fig.~\ref{transient plot}, we plot the time evolution to the steady state under strong decoherence and detection efficiency for both an initial state localized at the center (Fig.~\ref{transient plot}(a)) and biased towards one edge (Fig.~\ref{transient plot}(b)) of the cylinder graph. For the unbiased initial state, the system reaches the steady state rapidly. In contrast, the biased initial state, especially when $\eta$ and $p$ are high, causes the walker to become trapped in a metastable state where the probability of the walker remains high near the initialized edge. This trapping delays the transport of the walker across the bulk, requiring prolonged timescales to finally reach the steady state. From an experimental perspective, this demonstrates that preparing an unbiased, centered initial state is advantageous for accelerating the observation of the localization phenomenon.

\section{Localization Timescales under QSW}
\label{transient_dynamics}

Having fully characterized the properties of the steady state of QSW, we now study the localization time. To quantify how fast the system converges to its edge-localized steady state, we compute the trace distance between the time-evolved density matrix ($\rho(t)$) and the numerically determined steady state ($\rho^{ss}$) for a walker on the cylinder graph initially localized at its center. The trace distance is defined as:
\begin{equation}
D(\rho(t), \rho^{ss}) = \tfrac{1}{2} \, \mathrm{Tr}\!\left( \sqrt{(\rho(t) - \rho^{ss})^{\dagger} (\rho(t) - \rho^{ss})} \right).
\label{eq:tracedistance}
\end{equation}

This metric quantifies the distinguishability between the two states, with $D=1$ implying they are perfectly distinct and $D=0$ signifying that the system has fully converged to the steady state. The localization time, $\tau$, is defined as the time required for the trace distance $D(\rho(t), \rho^{ss})$ to decay below a threshold of $10^{-2}$.

\begin{figure}[h]
\centering
\includegraphics[width=0.5\textwidth]{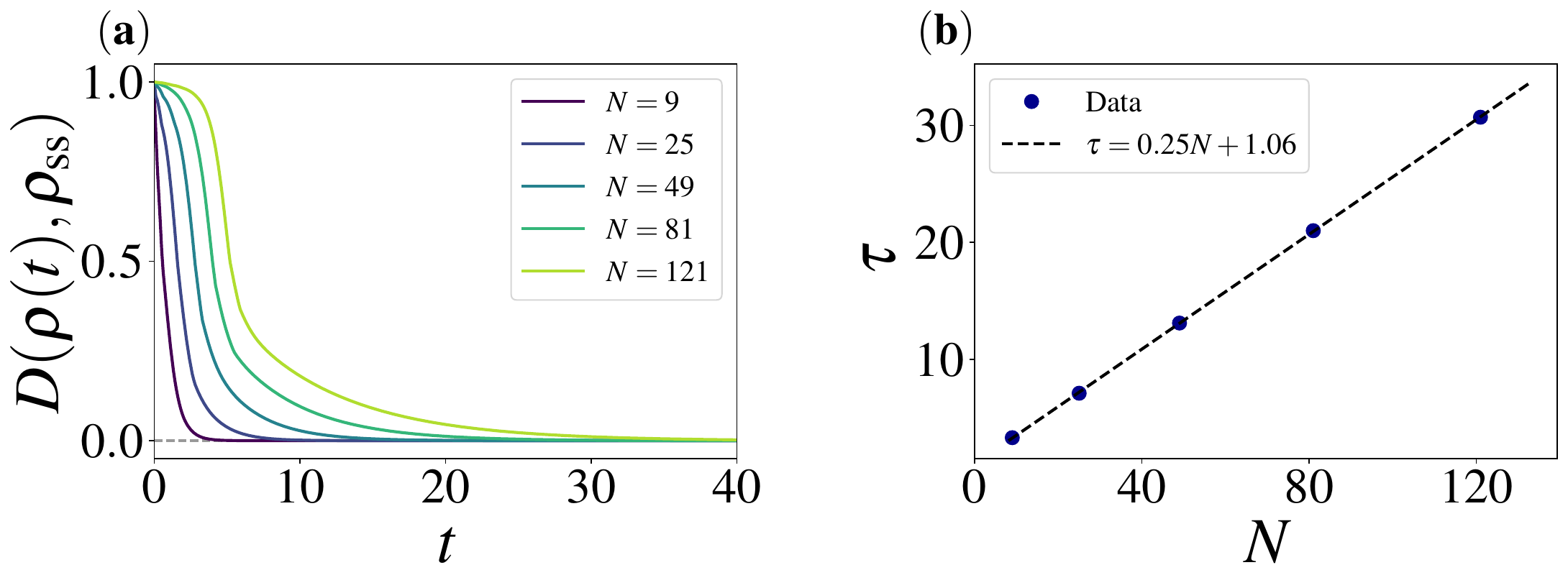}
\caption{Steady-state convergence and localization time scaling with graph size ($N$). (a) Time evolution of the trace distance for varying $N$ at fixed $p=0.5$ and $\eta=0.5$. (b) Dependence of the localization time $\tau$ on $N$. The results show that the time required for the quantum walker to reach steady state scales linearly with the graph size.}
\label{trace_distance_N}
\end{figure}

\begin{figure}[h]
\centering
\includegraphics[width=0.5\textwidth]{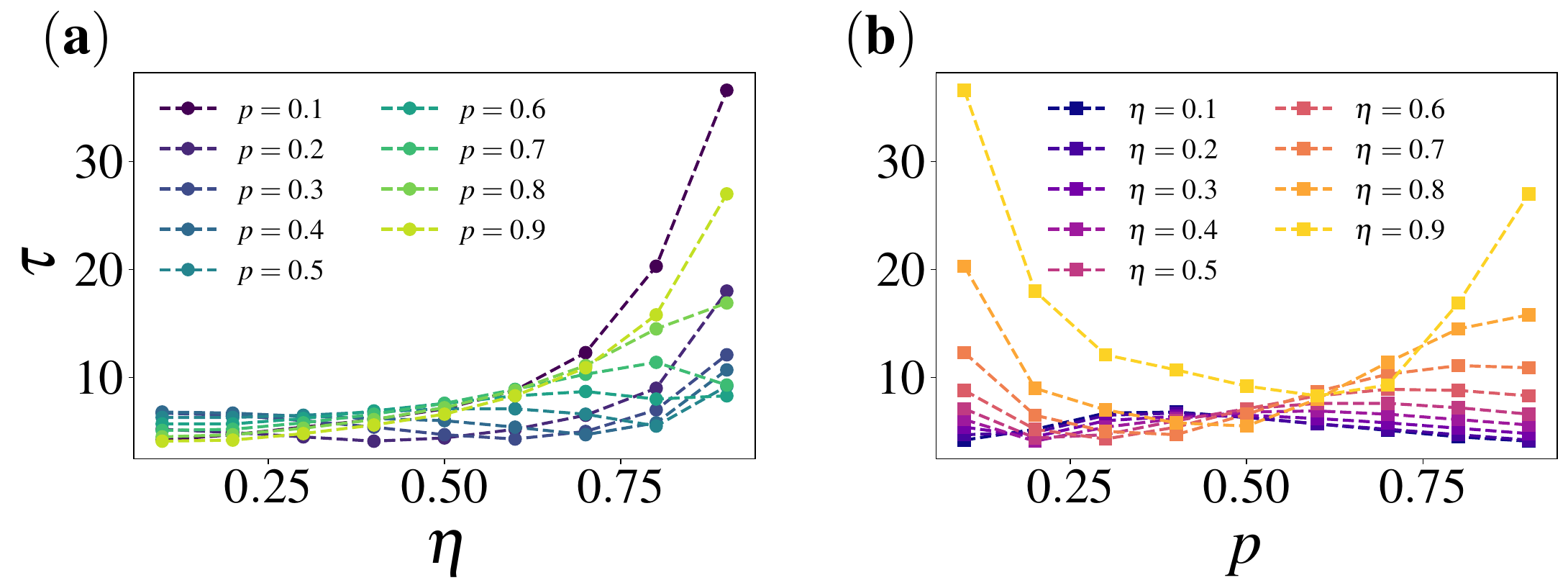}
\caption{Dependence of the localization time $\tau$ on the efficiency $\eta$ and the decoherence strength $p$. (a) $\tau$ versus $\eta$ for fixed values of $p$. (b) $\tau$ versus $p$ for fixed values of $\eta$. The data show that for large values of $\eta$, $\tau$ spikes at both weak and strong decoherence regimes. Optimal convergence to steady state is achieved at intermediate decoherence strength.}
\label{trace_distance}
\end{figure}

As shown for the cylinder graph in Fig.~\ref{trace_distance_N}(a), the trace distance exhibits a monotonic decay toward the steady state. As the graph size scales up, an initial transient plateau emerges, delaying the onset of decay and consequently increasing the overall localization time. To isolate this scaling from parameter induced effect, we fix $\eta =0.5$ and $p=0.5$. As demonstrated subsequently in Fig.~\ref{trace_distance}, these specific values lie within an optimal parameter region where the localization time does not change drastically with variations in either $p$ or $\eta$. Fig.~\ref{trace_distance_N}(b) shows that the localization time scales linearly with the total number of nodes, $\tau \propto N$, which arises naturally from the proportional expansion of the state space available to the quantum walker. 

To determine the fraction of trajectories that survive until the localized steady state is reached, we evaluate the survival probability, $S(\tau)$. This probability is defined as the trace of the unnormalized density matrix~\cite{bettmann2025quantum}, $S(t)=\text{Tr}(\tilde{\rho})$. Starting from the normalized conditional evolution in Eq.~\eqref{eq:15}, we obtain the dynamics of the unnormalized state by excluding the nonlinear normalization term. Evaluating the trace of $\dot{\tilde{\rho}}$, which corresponds to the negative of this nonlinear term, then gives:

\begin{equation}
    \frac{dS(t)}{dt}= -\eta\,p\sum_{k,j}\mathrm{Tr}(P_{kj}^\dagger P_{kj}\tilde{\rho})\ .
    \label{eq:S}
\end{equation}

Substituting $\tilde{\rho(t)}= S(t)\rho(t)$ and $\sum_j P_{kj}^\dagger P_{kj}= (D_j^2+D_j)|j\rangle\langle j|$ (Eq.~\eqref{eq:16}) into Eq.~\eqref{eq:S}, rearranging, and integrating gives:

\begin{equation}
    S(t)= \exp{\Bigl[-\eta p \int_0^t ds \sum_{j}(D_j^2+D_j)\rho_{jj}(s)\, \Bigr]}.
    \label{eq:integral}
\end{equation}

To extract the survival probability at the localization time, $S(\tau)$, and establish its scaling behavior with network size $N$, we separate the degree contributions from the edge nodes ($E$) and the bulk nodes ($B$) of the cylinder graph:

\begin{equation}
\begin{split}
    S(\tau) &= \exp\Bigl[-\eta p \int_0^\tau ds \Bigl(\sum_{j\in E}(D_E^2+D_E)\rho_{jj}(s) \\
    &\quad + \sum_{k\in B}(D_B^2+D_B)\rho_{kk}(s)\Bigr)\, \Bigr].
\end{split}
\label{eq:integral_1}
\end{equation}

Because the node probabilities must sum to unity, this integral is bounded. Substituting $\sum_{k\epsilon B}\rho_{kk}(s)=1-\sum_{j\epsilon E}\rho_{jj}(s)$ into Eq.~\eqref{eq:integral_1} yields:

\begin{equation}
\begin{split}
    S(\tau) &= \exp\Bigl[-\eta p (D_B^2+D_B) \tau \\ 
    &\quad +\eta p \left((D_B^2+D_B)-(D_E^2+D_E)\right)\int_0^\tau ds \sum_{j\in E}\rho_{jj}(s)\, \Bigr].
\end{split}
\label{eq:integral_2}
\end{equation}

Since $(D_B^2+D_B)>(D_E^2+D_E)$ and the integrated probability summation is strictly positive, the second term in the exponent evaluates to a positive quantity. Removing it decreases the total exponent, yielding the lower bound for the survival probability. Conversely, substituting $\sum_{j\epsilon E}\rho_{jj}(s)=1-\sum_{k\epsilon B}\rho_{kk}(s)$ into Eq.~\eqref{eq:integral_1} yields:

\begin{equation}
\begin{split}
    S(\tau) &= \exp\Bigl[-\eta p (D_E^2+D_E) \tau \\
    &\quad -\eta p \left((D_B^2+D_B)-(D_E^2+D_E)\right)\int_0^\tau ds \sum_{j\in B}\rho_{jj}(s)\, \Bigr].
\end{split}
\label{eq:integral_3}
\end{equation}

In this form, the second term in the exponent is a negative quantity. Removing it increases the total exponent, providing the upper bound for the survival probability. Together, these limits constraint $S(\tau)$ as:

\begin{equation}
    \exp{\Bigl[-\eta p (D_B^2+D_B) \tau \Bigr]} \leq S(\tau) \leq \exp{\Bigl[-\eta p (D_E^2+D_E) \tau \Bigr]}.
    \label{eq.C7}
\end{equation}

Both limits show that $S(\tau)$ must decay exponentially with respect to localization time, scaling as $S(\tau) \sim \exp{(-c\tau)}$. As established in Fig.~\ref{trace_distance_N}(b), the localization time itself scales linearly with the system size ($\tau\propto N$). Consequently, the survival fraction at the steady state is exponentially suppressed by the number of nodes, $S(\tau) \sim \exp{(-bN)}$. This implies that, as $N$ increases, the number of experimental repetitions required to observe the localized steady state grows exponentially.

To show how the dependence on $\eta$ and $p$ in Eq.~\eqref{eq:15} is coupled, we split the Liouvillian as a combination of an effective non-Hermitian Hamiltonian $H_{\mathrm{eff}}$, pure dissipative or pure measurement terms $\mathcal{L}_{M}$, and a nonlinear normalization term $\mathcal{L}_{N}$~\cite{liu2025lindbladian}:

\begin{equation}
\frac{d\rho}{dt}
=
-i H_{\mathrm{eff}}(\eta p) \rho+ i \rho H_{\mathrm{eff}}^{\dagger}(\eta p) 
+ \mathcal{L}_{M}(\rho)
+ \mathcal{L}_{N}(\rho),
\label{eq:R1}
\end{equation}

where \\
\begin{subequations}
\begin{equation}
    H_{\mathrm{eff}}(\eta p)= (1-p)H-\tfrac{i}{2}p\eta\sum_{k,j}P_{kj}^\dagger P_{kj} ,
    \label{eq:Heff}
\end{equation}

\begin{equation}
    \mathcal{L}_{M}(\rho)= p(1-\eta)\sum_{k,j}(P_{kj}\rho P_{kj}^\dagger-\tfrac{1}{2}\{P_{kj}^\dagger P_{kj}, \rho\}),
    \label{eq:LM}
\end{equation}

\begin{equation}
    \mathcal{L}_{N}(\rho)= p\eta\,\sum_{k,j}\mathrm{Tr}(P_{kj}^\dagger P_{kj}\rho)\,\rho.
    \label{eq:LN}
\end{equation}
\end{subequations}

As shown in Sect.~\ref{sec:QSW} (Fig.~\ref{bifurcation}), increasing both  $\eta$ and $p$ drives the system toward a steady state that is highly localized on the low-degree nodes. This edge-localization can be viewed as directly induced by the imaginary term in $H_{\mathrm{eff}}(\eta p)$. Conversely, the influence of $\mathcal{L}_{M}$ scales directly with $p(1-\eta)$, strengthening as $p$ increases and $\eta$ decreases, while $\mathcal{L}_{N}$ enforces trace preservation.

Fig.~\ref{trace_distance}(a) and (b) illustrate the influence of $p$ and $\eta$ on the localization time $\tau$. For smaller values of $\eta$ ($\eta\leq0.5$), the localization time remains small, indicating the localized steady state is attained faster compared to the large $\eta$ regime. However, as $\eta$ increases, the target steady state becomes highly localized on the low-degree nodes, delaying the steady state at both extremes of $p$. When $p$ is small, the unitary term $(1-p)H$ dominates (Eq.~\eqref{eq:Heff}), driving the quantum walker to delocalize across the network. However, the $\eta$-dependent imaginary term forces the steady state to be highly localized on the low-degree nodes; this competition causes a large delay in $\tau$. Conversely, at large $p$ and $\eta$, the edge localization becomes more prominent because the imaginary term dominates $H_{\mathrm{eff}}$. Simultaneously, the coherent transport from the unitary term becomes negligible, and the dissipative transport from $\mathcal{L}_{M}$ (Eq.~\eqref{eq:LM}) also diminishes, causing further delay and another spike in $\tau$. At intermediate decoherence ($p\approx0.4$ to $0.7$), the pure dissipative $\mathcal{L}_{M}$ term and $H_{\mathrm{eff}}(\eta p)$ are optimally tuned to balance the edge localization, resulting in an optimal convergence profile. 

\section{Steady State under Haken-Strobl Noise}
\label{sec:appendix_a}
In this appendix, we detail the derivation of the steady-state solution for Haken-Strobl noise. Starting from the NLME for Haken-Strobl noise by substituting Eq.~\eqref{HSeqn} into Eq.~\eqref{eq:1}, the anti-commutator term in the master equation can be significantly simplified. The entire expression can be reduced as follows:
\begin{equation}
    \sum_{k} \{P_{k}^\dagger P_{k}, \rho\} = \{\sum_{k}|k\rangle \langle k|,\rho\}= \{I,\rho\}= 2\rho.
    \label{eq:3}
\end{equation}
We now evaluate the Lindblad jump term, which describes the process of dephasing:
\begin{equation}
    \sum_{k} P_{k} \rho P_{k}^\dagger = \sum_{k} \rho_{kk}|k\rangle \langle k|=\mathcal P(\rho),
    \label{eq:4}
\end{equation}
where, $\mathcal{P}(\rho)$ is the projection onto diagonal elements of density matrix. Now the normalization term can be written as:
\begin{equation}
    \sum_k \mathrm{Tr}\left( P_k^\dagger P_k \, \rho \right)= \sum_k \mathrm{Tr}\left( |k\rangle \langle k|, \rho \right)= \sum_k \rho_{kk}=1.
    \label{eq:5}
\end{equation}
This shows that the nonlinearity of the NLME Eq.~\eqref{eq:1} is nullified because of the nature of the operators $\sum_k |k\rangle \langle k|= I$. Substituting Eq.~\eqref{eq:3}, Eq.~\eqref{eq:4} and Eq.~\eqref{eq:5} in Eq.~\eqref{eq:1} and rearranging yields:
\begin{equation}
\frac{d\rho}{dt}
= -i[H,\rho]
- \gamma (1-\eta)(\rho - \mathcal{P}(\rho)).
\label{eq:6}
\end{equation}
If the detectors are perfectly efficient ($\eta=1$), the prefactor ($1-\eta$) in Eq.~\eqref{eq:6} becomes zero, and the dissipative term vanishes entirely. The NLME then reduces to the purely unitary von Neumann equation, $\dot\rho=-i[H,\rho]$. In this limit, the system undergoes coherent evolution and does not relax to a steady state. Therefore, we restrict our steady-state analysis to the regime of postselection $\eta<1$.

We now consider the long-time behavior of the system, which is characterized by the steady-state density matrix, $\rho^{ss}$. Imposing the condition $\dot{\rho^{ss}}=0$  on Eq.~\eqref{eq:6} gives the following balance equation:
\begin{equation*}
 i[H,\rho^{ss}]=
- \gamma (1-\eta)(\rho^{ss} - \mathcal{P}(\rho^{ss})).
\end{equation*}
The structure of the steady-state equation in Eq.~\eqref{eq:7a} imposes a strong constraint on the form of $\rho^{ss}$. 

To determine the form of $\rho^{ss}$, we apply Hilbert Schmidt inner product argument to Eq.~\eqref{eq:7a}. Multiplying Eq.~\eqref{eq:7a} by $\rho^{ss}$ and taking the trace of both sides yields:
\begin{equation}
 \mathrm{Tr}(\rho^{ss}i[H,\rho^{ss}])=
- \gamma (1-\eta)\mathrm{Tr}(\rho^{ss}(\rho^{ss} - \mathcal{P}(\rho^{ss}))).
\label{eq1:8}
\end{equation}
Using the cyclic property of the trace, the left-hand side of Eq.~\eqref{eq1:8} evaluated to zero:

\begin{equation}
     \mathrm{Tr}(\rho^{ss}i[H,\rho^{ss}])= 0.
     \label{eq1:9}
\end{equation}

Consequently, the right-hand side of Eq.~\eqref{eq1:8} must also vanish:

\begin{equation}
    \mathrm{Tr}(\rho^{ss}(\rho^{ss} - \mathcal{P}(\rho^{ss})))=\sum_{jk}\rho^{ss}_{jk}(\rho^{ss}-\mathcal{P}(\rho^{ss}))_{kj}=0.
    \label{eq1:10}
\end{equation}

For $j=k$, the diagonal elements of $\rho^{ss}-\mathcal{P}(\rho^{ss})=0$. For off-diagonal elements ($j \neq k$), $\mathcal{P}(\rho^{ss})$ has no entries. The sum therefore reduces to $\sum_{j\neq k}\rho^{ss}_{jk}\rho^{ss}_{kj}$. Applying the Hermitian condition $\rho^{ss}_{kj}=\rho^{ss}_{jk}*$, we obtain:

\begin{equation}
\sum_{j\neq k}|\rho^{ss}_{jk}|^2=0.
    \label{eq1:11}
\end{equation}

Since this is a sum of non-negative absolute squares, every off-diagonal entry must be zero. This implies that $\rho_{ss}$ is a purely diagonal matrix, satisfying $\rho^{ss}=\mathcal{P}(\rho^{ss})$. Substituting this condition into Eq.~\eqref{eq:7a} gives $[H, \rho^{ss}] = 0$. We now use this commutation relation to determine the diagonal elements. Let $\rho^{ss} = \sum_k \rho^{ss}_{kk}|k\rangle\langle k|$. The matrix elements of the commutator in the site basis are:

\begin{equation}
    [H, \rho^{ss}]_{jk} = H_{jk}\rho^{ss}_{kk} - \rho^{ss}_{jj}H_{jk} = H_{jk}(\rho^{ss}_{kk} - \rho^{ss}_{jj}).
    \label{eq:10}
\end{equation}

For the commutator in Eq.~\eqref{eq:10} to be zero, we must have $H_{jk}(\rho^{ss}_{kk} - \rho^{ss}_{jj})=0$ for all pairs of sites $(j, k)$. For any physical network where the Hamiltonian connects the sites (i.e., for any two nodes $j$ and $k$ connected by an edge, $H_{jk} \neq 0$), this forces their populations to be equal: $\rho^{ss}_{jj} = \rho^{ss}_{kk}$. For any connected graph, this condition propagates across the entire graph, implying that all diagonal elements of the steady-state density matrix must be identical. Given the trace constraint, $\mathrm{Tr}(\rho^{ss}) = \sum_k \rho^{ss}_{kk} = 1$, and that all $N$ diagonal elements are equal, each must be $\frac{1}{N}$. Thus, for $\eta<1$, this leads to the unique steady state  $\rho^{ss}=\frac{\mathbb{I}}{N}$.

\bibliography{reference}

\end{document}